\documentclass[aps,prl,twocolumn,reprint,secnumarabic]{revtex4-2}
\usepackage{latexsym}
\usepackage{graphicx}
\usepackage{times,psfrag,subfigure}
\usepackage{enumerate}
\usepackage{amsmath}
\usepackage{dsfont}
\usepackage{dcolumn}
\usepackage{bm,bbm}
\usepackage{color}
\usepackage{latexsym,amsmath,amssymb,bm,euscript}
\usepackage{dsfont}
\usepackage{textcomp}
\usepackage{tabularx}
\usepackage{setspace}
\usepackage{ctable}
\usepackage{sidecap}
\usepackage{placeins}
\usepackage{threeparttable}
\usepackage{multirow}
\usepackage{comment}
\usepackage{ulem}
\usepackage{amssymb,bm}
\usepackage{hyperref}
\usepackage{physics}
\usepackage{cancel}
\usepackage{ulem}
\hypersetup{colorlinks=true,
            citecolor=blue,
            linkcolor=blue,
            urlcolor=blue,
            bookmarksnumbered=true}

\let \oldbm \bm
\renewcommand{\vec}[1]{\oldbm{#1}}

\def\bk{{\vec k}}
\def\bu{{\vec u}}

\def\bxi{{\vec \xi}}

\def\ba{{\vec a}}
\def\bz{{\vec z}}
\def\bK{{\vec K}}
\def\bq{{\vec q}}
\def\bQ{{\vec Q}}

\def\bR{{\vec R}}
\def\bG{{\vec G}}

\def\bp{{\vec p}}

\def\br{{\vec r}}

\def\tr{\mathop{\mathrm{tr}}}
\def\Tr{\mathop{\mathrm{Tr}}}

\def\BZ{\mathrm{BZ}}

\def\H{\mathcal{H}}

\def\M{\mathcal{M}}

\def\SU{\rm{SU}}

\def\A{\mathcal{A}}

\begin{document}

\title{Microsopic Theory of Spin Polarons in Chern Ferromagnets}

\author{Qiang Gao}
\affiliation{Department of Physics, Harvard University, Cambridge, MA 02138, USA}
\author{Eslam Khalaf}
\affiliation{Department of Physics, Harvard University, Cambridge, MA 02138, USA}

\date{\today}

\begin{abstract}
Charged excitations in Chern ferromagnets play a central role in determining  the nature of doped phases in moiré heterostructures.
We develop a microscopic theory of charged excitations in an SU(2) Chern ferromagnet and obtain \textit{closed-form wavefunctions} for a hierarchy of charge-$e$ spin polaron states binding an arbitrary number of spin flips.
In an ideal Chern-$1$ band with a normal-ordered contact interaction, we show that these polarons are \textit{exact eigenstates} 
of the Hamiltonian with the same energy as single-hole excitations. Away from this ideal limit, we promote these states to a variational family by introducing a single size parameter and (in dispersive Chern bands) a geometry-informed single-particle dressing. Our momentum-space wavefunctions admit two equivalent representations: a ratio of jastrow factors of Weierstrass functions of relative momenta or an antisymmetrized geminal product of particle-hole wavefunctions. The latter 
enables efficient evaluation of overlaps and expectation values for large system sizes and large number of spin flips.
Benchmarking in the lowest Landau level, the single-spin-flip ansatz achieves $\gtrsim 99\%$ overlap with exact diagonalization and accurately captures binding energies, while the multi-spin-flip energies interpolate smoothly toward the large-texture (skyrmion) regime. For Chern bands with tunable quantum geometry, we find that interaction-generated single particle dispersion quickly destabilizes the spin polarons once quantum geometry becomes sufficiently non-uniform. When such dispersion is suppressed, however, the bound states persist deeper into the non-uniform regime, with the binding energy slowly decreasing and the bound state becoming larger as the quantum geometry becomes more concentrated.  Our results provide a microscopic foundation for analyzing doped Chern ferromagnets in moiré platforms and lay the groundwork for variational wavefunctions of multi-polaron excitations and phases.
\end{abstract}

\maketitle

\textit{Introduction.}---Nearly flat Chern bands, where electron interaction is the dominant energy scale, are now routinely realized in moir\'e heterostructures. Various correlated phases have been observed at partial filling of these bands, including correlated insulators, superconductors, and quantum anomalous Hall states 
\cite{CaoIns,CaoSC,Yankowitz2019,Lu2019,Sharpe2019,Serlin2020,Tang2020,Regan2020, KimTDBG, PabloTDBG, YankowitzTDBG, YankowitzMonoBi1, xu2023observation, zeng2023thermodynamic, cai2023signatures, park2023observation, YoungTrilayer2021, YoungRhombohedral2025, JuRhombohedralQAH}. At integer fillings, strong correlations often stabilize flavor ferromagnets --- multi-flavor generalizations of quantum Hall ferromagnets (QHFM) --- where a subset of spin/valley/Chern flavors is fully filled
\cite{YahuiChernBands,Bultinck2020,Khalaf2021,Lian2021,ledwith2021strong}.
The nature of \textit{charged} excitations above these ferromagnets is central: it sets transport gaps, controls compressibility, and determines the nature of the state obtained upon doping. In conventional QHFMs, it is known that skyrmions carry a charge and are the cheapest charge excitations upon doping \cite{Sondhi1993,Moon1995,Fertig1994}. In a Chern ferromagnet, topology also dictates that skyrmions are charged but their energetic competition vs single particle excitation is less understood and depends on the detailed quantum geometry of the band and its dispersion. These lattice scale features are challenging to incorporate within the standard description of skyrmions as smooth slowly varying texture.

To address this, earlier work either considered the extreme opposite limit of very small (baby) skyrmion with a single spin flip \cite{KhalafBabySkyrmions, schindlerTrionsTwistedBilayer2022} or performed numerical Hartree-Fock calculations on specific models \cite{KwanSkyrmion}. These approaches fail to reveal the general analytic structure of these excitations, their dependence on the dispersion and quantum geometry of the parent band, and how they evolve from the single spin flip limit to that of large smooth skyrmion textures. A microscopic understanding of charged excitations in Chern bands that accounts for their momentum space structure is thus needed as a necessary step for a theory of doped Chern ferromagnets.

Here we develop such microscopic theory for spin polarons that accounts for momentum space quantum geometry and dispersion and interpolate between few-spin bound states and large-$n$ skyrmion-like textures. Our starting point is an analytically tractable limit---an ideal Chern-$1$ band with a contact interaction---where we construct exact charge-$\pm e$ eigenstates that bind an arbitrary number of spin flips $n$.
We then show that the same wavefunctions become excellent variational ans\"atze for realistic interactions/band geometry, with a single variational parameter characterizing its effective size. We show that such wavefunctions have more than 99\% overlap with numerical wavefunctions in the single spin-flip case and reproduce the field theory skyrmion energies for QHFMs in the limit of large $n$. For an ideal Chern 1 band with tunable quantum geometry, we show that the size and binding energy depend sensitively on Berry curvature distribution; with more concentrated quantum geometry leading to larger bound states and smaller binding energy within a fixed spin sector. These wavefunctions set the stage for the construction of variational wavefunctions for phases of doped spin polarons, that accurately capture the energetics of binding between electron/hole and spin flips.

\textit{Projected interaction and Chern ferromagnets.}---
We consider two nearly flat Chern bands labeled by $\sigma=\uparrow,\downarrow$ with Chern number $C=1$ with ${\rm SU}(2)$ spin rotation symmetry.
Projecting a density--density interaction to these bands gives
\begin{equation}
H=\sum_{\bk,\sigma} \epsilon(\bk) c_{\sigma,\bk}^\dagger c_{\sigma,\bk} + \frac{1}{2A}\sum_{\bq} V(\bq)\,:\rho_{-\bq}\,\rho_{\bq}:,
\label{eq:Hproj}
\end{equation}
with area $A$ and projected density
\begin{equation}
\rho_{\bq}=\sum_{\bk,\sigma}\lambda_{\bq}(\bk)\,
c^{\dagger}_{\bk,\sigma}c_{\bk+\bq,\sigma},\qquad
\lambda_{\bq}(\bk)=\langle u_{\bk}\vert u_{\bk+\bq}\rangle .
\label{eq:rho}
\end{equation}
 At filling $\nu=1$, the fully polarized state, e.g. $\ket{\downarrow}$, is an exact eigenstate of~\eqref{eq:Hproj}. Furthermore, this state is guaranteed to be a ground state for ideal flat bands with contact interaction as we will show later. As a result, we expect $\ket{\downarrow}$ to remain the ground state for a range of parameters away from this ideal limit, consistent with findings from numerical studies in moir\'e systems \cite{Bultinck2020,Lian2021, TBGMonteCarlo, TBGDMRG, TBGDMRGStrain}, which we will assume in the following.
We focus on charge-$e$ excitations built by adding $n$ electrons and $n{+}1$ holes to $\ket{\downarrow}$, corresponding to $n$ spin flips. For a system with $N$ unit cells, these are states with $S_z = \frac{N-1}{2} - n$, whose total spin can take values $S_{\rm total} = \frac{N-1}{2} - n + r$ with $r = 0,1,\dots,n$. Any state labeled by $(n,r)$ with $r \neq 0$ can be generated by applying the spin lowering $r$ times to a state with $(n-r,0)$, which amounts to attaching $\bq = 0$ Goldstone modes. Thus, we can focus on states with $r = 0$ within a given $n$ sector (see SM for details \cite{SM}). Note that the minimum energy as a function of $n$ can only decrease, since we can always attach a goldstone mode with finite but arbitrarily small $|\bq|$ to the minimum energy state in the $n$ sector to get a state with arbitrarily close energy in the $n+1$ sector. \footnote{Adding the $\bq = 0$ Goldstone mode amounts to rotating the parent state. This gives rise to a state where $S_z$ is decreased by 1 but $S_{\rm total}$ is unchanged. However, for non-zero but small $\bq$, this state has different total spin $S_{\rm total}$.} 

\begin{figure*}[t]
\centering
\includegraphics[width=0.88\textwidth]{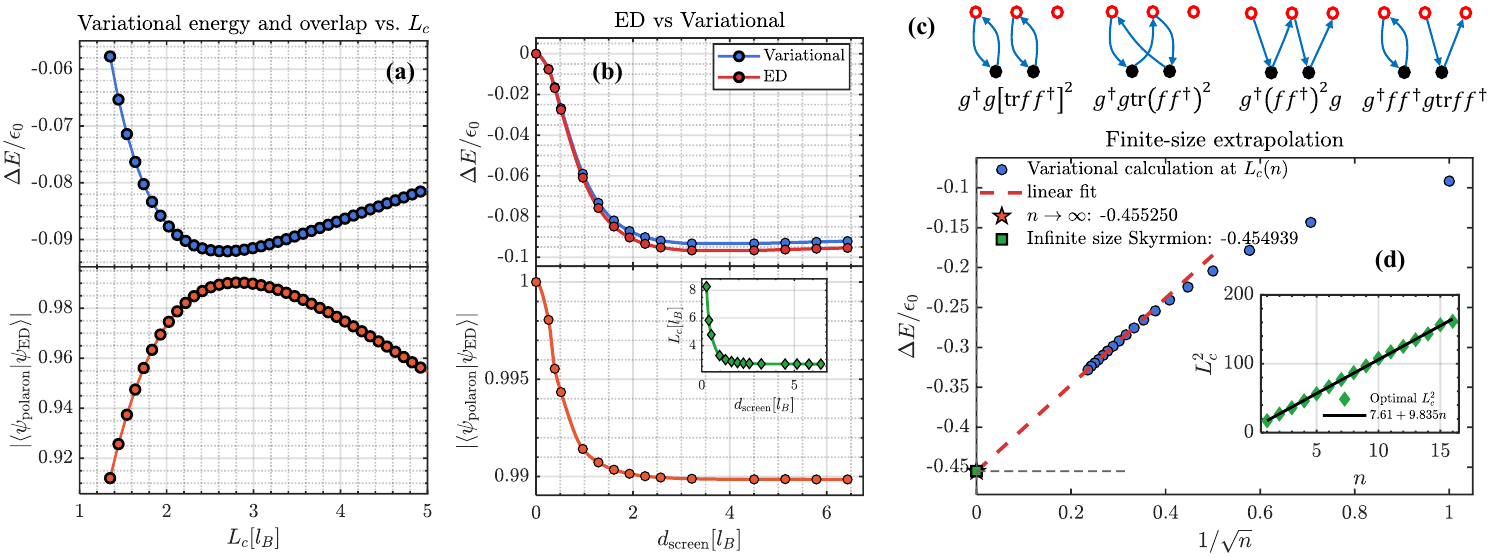}
\caption{Lowest Landau level (LLL) benchmark. (a,b) For a single-spin-flip ($n{=}1$) polaron: binding energy $\Delta E$ relative to a single hole (top) and overlap with the exact numerical wavefunction (bottom) versus (a) inverse size parameter $L_c$ with fixed $d=10$ and (b) screening length $d$ for screened Coulomb interaction with optimized $L_c$. The system size for (a,b) is $19\times19$. (c) Diagrammatics for the $n=2$ normalization (and related) integrals in the geminal representation.
Closed loops represent traces of powers of $F$, while open loops represent $g^{\dagger}F^{\ell}\tilde g$; see Eq.~\eqref{eq:AB}. (d) Variational energy for a polaron with $n$ spin flips versus $1/\sqrt{n}$ at large screening distance $d=10$ with $L_c$ fixed to be $L_c^2=7.61+9.835n$, illustrating the evolution toward the large-texture regime. The system size is $23\times 23$. The $L_c(n)$ is concluded from optimizations of a smaller system ($15\times15$) shown in the inset. In all plots, we have $l_B^2=4\pi/3\sqrt{3}$.}
\label{fig:variational_polaron}
\end{figure*}

\textit{LLL with contact interactions and exact zero modes.}--- It is known that in a QHFM in the LLL, skyrmions are favored to single particle excitations for Coulomb interaction \cite{Moon1995, Sondhi1993}, but have the same energy (independent of the skyrmion size) for contact interaction \cite{Fertig1994}. This suggests the latter is a good starting point to construct exact eigenstates with charge $+e$ and $n$ spin flips that has the same energy but are distinct from Goldstone-dressed holes. Once we deviate from this limit, we will see that these become excellent variational states whose energy is below a single hole, signifying the formation of true bound states.

To construct these states, we work with the projected operators
\begin{equation}
    c_{\br, \sigma} = \sum_\alpha \phi_\alpha(\br) c_{\alpha,\sigma}, \quad \{c_{\br, \sigma}^\dagger, c_{\br', \sigma'}\} = \delta_{\sigma,\sigma'} P^{\rm LLL}(\br, \br')
\end{equation}
where $P^{\rm LLL}(\br, \br') = \frac{1}{2\pi} e^{\frac{1}{4} (2r_1 \bar r_2 - |r_1|^2 - |r_2|^2)}$ is the LLL projector (with $l_B = 1$) and $\phi_\alpha(\br)$ is any orthonormal basis of LLL states. To make the connection to ideal bands, we will use Bloch basis which are eigenstates of commuting lattice magnetic translation (see S.M.~\cite{SM} for review).The normal ordered contact interaction takes the simple form:
\begin{equation}
H_{0}= \int \dd^2 r\, \hat M^{\dagger}(\br)\hat M(\br),\qquad
\hat M(\br)=c_{\br,\downarrow}c_{\br,\uparrow},
\label{eq:contact}
\end{equation}
Note that: (i) $H_0$ is a sum of positive semi-definite terms, (ii) $\hat M(\br) \ket{\downarrow} = 0$ which implies $\ket{\downarrow}$ is an exact ground state, (ii) any state $\hat O\ket{\downarrow}$ with $[\hat O,\hat M(\br)]=0$ is a zero mode, and as a result (iii) a single hole on top of $\ket{\downarrow}$ is a zero energy state \footnote{for a density-density interaction, which differs from the normal-ordered interaction by a constant, a hole costs a constant positive energy} and so is any state with any number of holes \footnote{since the holes in the same band cannot interact via contact interaction due to Pauli blocking}. On the other hand, it is non-trivial to construct zero modes that involve electron operators $c_{\br,\uparrow}^\dagger$ which generally do not commute with $\hat M(\br)$ \footnote{as discussed earlier, we are excluding states formed by a combination of a hole and $n$ goldstone modes which satisfy this condition trivially} .

For charge $+e$ and one flipped spin, the unique (up to gauge choices) nontrivial operator satisfying  $[\hat O,\hat M(\br)]=0$ is \cite{SM}
\begin{equation}
\ket{\Psi^{(1)}_{\bxi}}=\hat S_{\bxi}\,c_{\bxi,\downarrow}\ket{\downarrow},\qquad
\hat S_{\bxi}=\int \dd^2 \bz \,\frac{c^{\dagger}_{\bz,\uparrow}c_{\bz,\downarrow}}{z-\xi},
\label{eq:polaronop}
\end{equation}
where the unbolded letters indicate complex number corresponding to a given vector, e.g. $\xi = \xi_x + i \xi_y$. To verify that $\hat O$ commutes with $\hat M(\br)$, we use the following important identity (a proof is provided in SM \cite{SM}): for $\psi_\xi(z) = \frac{f(z)}{z - \xi} e^{-\frac{1}{4}|z|^2}$ where $f(z)$ is holomorphic, the LLL projection yields $[P^{\rm LLL} \psi_\xi](z) = \frac{e^{-\frac{1}{4}|z|^2}}{z - \xi}[f(z) - f(\xi)]$. Note that the projection has removed the pole at $z = \xi$. More generally,
\begin{equation}
\ket{\Psi^{(n)}_{\bxi}}=\big(\hat S_{\bxi}\big)^n\,c_{\bxi,\downarrow}\ket{\downarrow}
\label{eq:polaronopn}
\end{equation}
is an \textit{exact} zero-energy eigenstate of~\eqref{eq:contact} for every integer $n\ge 0$, describing a charge-$+e$ excitation with $n$ spin flips (a ``spin polaron'').

On a torus, writing momenta as complex numbers, $k=k_x+ik_y$, and using the Weierstrass $\sigma$ function \cite{HaldaneRezayi1985}, the corresponding momentum-space wavefunction has a compact universal form.
For $n$ spin flips (\,$n$ electrons at $\{\bk^e_{b}\}_{b=1}^{n}$ and $n{+}1$ holes at $\{\bk^h_{a}\}_{a=1}^{n+1}$), we have \cite{SM}
\begin{multline}
\psi^{\rm LLL}_{\bxi}(\{\bk^e\},\{\bk^h\})
=\varphi^{\mathrm{LLL}}_{\bQ}(\bxi)
e^{-\tfrac14\big(\sum_{a}|k^h_{a}|^2-\sum_{b}|k^e_{b}|^2-|Q|^2\big)} \\
\times
\frac{\prod_{b<b'}\sigma(k^e_{b}-k^e_{{b'}})\,\prod_{a<a'}\sigma(k^h_{a}-k^h_{{a'}})}{\prod_{a,b}\sigma(k^e_{b}-k^h_{a})},
\label{eq:psi_general}
\end{multline}
where $\bQ=\sum_a \bk^h_{a}-\sum_b \bk^e_{b}$ is the total momentum and $\varphi^{\mathrm{LLL}}_{\bQ}(\bxi)$ is an LLL Bloch wavefunction evaluated at $\bxi$. We can construct a momentum eigenstate as
\begin{equation}\label{momentum_eigenstate_LLL}
   \psi^{\rm LLL}_{\bk} = \int_{\rm UC} d^2 \bxi [\varphi^{\mathrm{LLL}}_{\bQ}(\bxi)]^* \psi^{\rm LLL}_{\bxi} \propto \delta([\bk - \bQ])
\end{equation}
where $[\bk]$ denotes the part of $\bk$ in the first BZ.

There is an equivalent form of this wavefunction that elucidates its structure:
\begin{gather}
    \psi^{\rm LLL}_\bxi(\{\bk^e\},\{\bk^h\}) = {\mathcal A} g_\bxi(\bk^h_{n+1}) \prod_{l=1}^n f_\bxi(\bk^h_{l}, \bk^e_l),  \nonumber\\ g_\bxi(\bk) = \phi^{\rm LLL}_\bk(\bxi), \nonumber \\ f_\bxi(\bk^h, \bk^e) = \frac{g_\bxi(\bk^h)}{g_\bxi(\bk^e)} [\zeta(k^h - i \xi) - \zeta(k^h - k^e)]
    \label{PsiNGeminal}
\end{gather}
where $\zeta(k)$ is the periodic Weierstrass $\zeta$ function defined as $\zeta(k) = \frac{\sigma'(k)}{\sigma(k)} - \frac{1}{2} \bar k$ and $\mathcal{A}$ antisymmetrizes over hole and electron momenta (details in S.M.~\cite{SM}).
This form resembles an antisymmetrized geminal power (AGP) state familiar in quantum chemistry \cite{Coleman1965,BlaizotRipka} and can be understood as a number projected excitonic BCS wavefunction where $f$ plays the role of the electron-hole pair wavefunction.

We note from the expression (\ref{eq:psi_general}) and (\ref{PsiNGeminal}) that the wavefunction has a pole when any electron momentum $\bk^e_b$ equals a hole momentum $\bk^h_a$, which leads a logarithmic divergence in the normalization integral for the wavefuntion. This pole arises from the logarithmic divergence of the integral in (\ref{eq:polaronop}) for large $\bz$ when $\bk^e_a = \bk^h_b$ for some $a$ and $b$. In any finite system, this divergence will be cutoff by the system size, which introduces an IR regulator $L_c \sim L$ for the $\bz$ integeral in (\ref{eq:polaronop}). The resulting small momentum cutoff is of the order of the momentum grid spacing $\Delta \bk \sim \frac{1}{L} \sim \frac{1}{L_c}$ which removes the pole at $\bk^e_b = \bk^h_a$.
We can introduce such cutoff in the final form of the momentum space wavefunction (\ref{eq:psi_general}) and (\ref{PsiNGeminal}) as follows
\begin{equation}
    \frac{1}{\sigma(\bk)} \mapsto  \frac{F_{L_c}(\bk)}{\sigma(\bk)}, \quad f_\xi(\bk^h, \bk^e) \mapsto f_\xi(\bk^h, \bk^e) F_{L_c}(\bk^h - \bk^e)
\end{equation}
with $F_{L_c}(\bk) = 1 - e^{-L_c^2 \bk^2}$.

The IR divergence of the integral in the polaron wavefunctions reflects the fact that for contact interactions, polarons are not true bound states as they have the same energy as a single hole. Instead, they are critical states whose size diverges with the system size. As we will see later, their size becomes finite when we consider interactions beyond the contact limit where they become proper bound states.

\textit{Variational polarons beyond the contact limit.---} Away from the contact limit, the states (\ref{eq:polaronopn}) are no longer exact eigenstates. For finite but sufficiently short-range interactions, we expect that any polaron bound state should be well approximated by the wavefunction (\ref{eq:psi_general}) in a finite system. However, as the system size grows beyond the spatial extent of the bound state, the overlap between the two must decrease and ultimately vanish in the thermodynamic limit, since one is a bound state while the other is extended. This motivates promoting $L_c$ from a quantity set by the system size to a variational parameter representing the size of the bound state. Doing so allows us to exploit the short-distance structure of the states (\ref{eq:psi_general}), which encodes the universal attraction between a charge and the dipolar spin fluctuations induced by band topology \cite{KhalafBabySkyrmions}.

In Fig.~\ref{fig:variational_polaron}(a), we show the binding energy $\Delta E$ ---defined as the energy of the variational ansatz relative to the single-hole energy --- as a function of the inverse size parameter $L_c$ for Coulomb interactions. A clear minimum at finite $L_c$ signals the existence of a bound state. To benchmark this variational approach, we compare our results to exact diagonalization of the three-particle Hamiltonian following Ref.~\cite{KhalafBabySkyrmions}. Figure~\ref{fig:variational_polaron}(b) displays both the overlap between the variational and numerically exact wavefunctions and the corresponding energies, for a screened Coulomb interaction as a function of the gate distance. The variational states achieve an overlap exceeding 99\%
 with the exact eigenstates, approaching 100\% in the contact limit $d \rightarrow 0$. The variational energies likewise converge to the exact values as the screening length is reduced.

\textit{Generalization to ideal Chern bands}--- We will now show how to generalize the construction to ideal bands, first by constructing an exact polaron zero mode for flat bands $\epsilon(\bk) = 0$ with contact interactions, then by showing how this can be used to construct variational states for more general interactions that also allow for finite band dispersion. An ideal $C = 1$ band is equivalent to the LLL of a Dirac particle in a periodic magnetic field up to a momentum independent spinor, with the wavefunction given explicitly by~\cite{Ledwith2019,LandauLevelWang, LedwithVishwanathParker22}
\begin{equation}
    \phi^a_\bk(\br) = \mathcal{N}(\bk)e^{\tfrac 14 Q(\br)} \phi^{\rm LLL}_\bk(\br) \chi^a(\br). 
    \label{eq:AC_wavefunction}
\end{equation}
Here, $\phi^{\rm LLL}_\bk(\br)$ are the Bloch states for the LLL in uniform field, $Q(\br)$ is a periodic function, $\mathcal{N}(\bk)\equiv \exp(\beta(\bk)/2+l_B^2\bk^2/4)$ is a normalization factor, and $\chi^a(\br)$ is a spinor in orbital indices satisfying $\| \chi(\br)\| = 1$. This allows us to introduce projected operators, as in the LLL case, such that $\rho(\br) = \sum_\sigma c^\dagger_{\br,\sigma} c_{\br,\sigma}$ and $\{c_{\br}^\dagger, c_{\br'}\} = P(\br, \br')$ where $P$ is the band projector. The functions $Q(\br)$ and $\beta(\bk)$ are related by an integral transform \cite{YanAnyonDispersion} (see detailed discussions in \cite{SM}) and encode the variations of the non-uniform magnetic field and Berry curvature via
\begin{equation}
    \delta B(\br) = -\nabla^2 Q(\br), \qquad \delta \Omega(\bk) = \tfrac 12\nabla_\bk^2 \beta(\bk)
    \label{deltaOmega}
\end{equation}
where $\delta B(\br)$ and $\delta \Omega(\bk)$ are defined by subtracting their average value so that they have vanishing averages over the unit cell or Brillouin zone, respectively.

To construct the zero mode, we note that we can write the Hamiltonian for a normal ordered contact interaction in the form (\ref{eq:contact}) with the replacement $\hat M \mapsto T^{-1} \hat M T$ where $T$ is the invertible but non-unitary transformation $T = e^{\frac{1}{2} \sum_{\bk,\sigma} \beta(\bk) c_{\bk,\sigma}^\dagger c_{\bk,\sigma}}$ and $\hat M$ is the LLL operator \cite{schleith2025anyon,YanAnyonDispersion}. This means that we can map the zero mode in the LLL problem to the zero mode in an ideal Chern 1 band via $|\Psi_0 \rangle = T|\Psi_0^{\rm LLL} \rangle$, yielding the first-quantized wavefunction
\begin{multline}
    \psi_{\bxi}^{\rm Ideal}(\{\bk^e_l\},\{\bk^h_l\}) = e^{-\frac{1}{2} [\sum_{l=1}^{n+1} \beta(\bk^h_l) - \sum_{l=1}^{n} \beta(\bk^e_l)]} \\ \times \psi_{\bxi}^{\rm LLL}(\{\bk^e_l\},\{\bk^h_l\})
\end{multline}
where $\psi^{\rm LLL}$ is given by the equivalent expressions (\ref{eq:psi_general}) and (\ref{PsiNGeminal}). We can construct the momentum eigenstates in analogy to Eq.~\eqref{momentum_eigenstate_LLL}.

In a Chern band, dispersion is generally allowed by symmetry and it is usually generated by interaction even if the original Hamiltonian has no single particle term. To account for the dispersion, we can include a one-body dressing that biases weight toward the minima of the interaction-generated quasiparticle dispersion. Since the interaction-generated dispersion has a similar shape to $\beta(\bk)$, a convenient choice is
\begin{equation}
\psi_{\mathrm{var}}\propto
\exp\!\Big[-\tfrac{1+\eta_h}{2}\sum_{a}\beta(\bk^h_{a})+\tfrac{1-\eta_e}{2}\sum_{b}\beta(\bk^e_{b})\Big]
\psi^{\mathrm{LLL}}_{\bxi,L_c},
\label{eq:var}
\end{equation}
with variational parameters $L_c$, $\bxi$, and $\eta_{e,h}$. \footnote{Strictly speaking, we should construct the variational wavefunctions at fixed momentum and find the minimum energy as a function of momentum. This will provide information about the polaron dispersion. However, since the momentum wavefunctions do not admit a simple geminal form, we take instead the polaron position as a variational parameter and minimize over it to get the minimum polaron energy. This is expected to provide a good variational estimates since polaron kinetic energy is expected to be small}

\textit{Diagrammatic evaluation for multiple spin flips.}---
The main technical obstacle for $n>1$ is the high-dimensional integrals in expectation values, e.g.
$E_{\mathrm{var}}=\bra{\psi}H\ket{\psi}/\bra{\psi}\psi\rangle$,
whose brute-force evaluation is hindered by the near-singular structure inherited from the exact wavefunctions.
A crucial simplification is obtained by using the \textit{geminal} representation of the $n$-spin-flip wavefunction (\ref{PsiNGeminal}).
For any function of that form (irrespective of the specific form of $f$ and $g$) and for a given $\bk$-grid, we can think of $g$ and $f$ as a vector and a matrix with $g_i\equiv g(\bk^h_i)$ and $f_{ij}\equiv f(\bk^h_i,\bk^e_j)$.

Overlaps and the auxiliary overlaps needed for operator insertions, say $\langle\psi|\tilde\psi\rangle$, reduce to a finite sum of diagrams built from a matrix
$F\equiv \tilde f f^{\dagger}$ and vectors $g,\tilde g$, through the basic building blocks
\begin{equation}
A_{\ell}\equiv \Tr F^{\ell},\qquad B_{\ell}\equiv g^{\dagger}F^{\ell}\tilde g,\qquad (\ell\le n).
\label{eq:AB}
\end{equation}
For $n=2$, the four contributing topologies are shown in Fig.~\ref{fig:variational_polaron}(c).
Each closed loop with $\ell$ electron nodes contributes $A_{\ell}$, while an open loop contributes $B_{\ell}$.
For general $n$, the closed-loop class is in one-to-one correspondence with integer partitions of $n$, and the full overlap can be written as an explicit sum over partitions with computable sign and combinatorial weights (see S.M.~\cite{SM}).
Once $\{A_{\ell},B_{\ell}\}$ are computed, evaluating any individual diagram is $\mathcal{O}(1)$, and in practice the method is efficient up to $n\sim 20$ on the system sizes relevant for variational energies. Expectation values of operators can be reduced to the same overlap class by introducing a source parameter $\varepsilon$ in $g$ and $f$ and differentiating overlaps with respect to $\varepsilon$ (see S.M.~\cite{SM}).
This yields a deterministic alternative to Monte Carlo for singular integrals and allows systematic study of multi-spin-flip polarons.

\textit{Multi-spin-flip polarons and approach to the skyrmion limit.}---
Fig.~\ref{fig:variational_polaron}(d) shows the variational energy for polarons with multiple spin flips for large screening distance $d \sim 10$. To connect with the skyrmion limit at large $n$, we note that a skyrmion with radius $R$ typically involves $R^2$ spin flips and its Coulomb energy goes down to the asymptotic minimal value of $4\pi \rho$ as $1/R \sim 1/\sqrt{n}$. We also find numerically that the optimal cut-off $L_c$ for LLL follows a simple relation with $n$ as $L_c^2=C_0+C_1n$ with the coefficients $C_i$ depending on the interactions (see the inset in Fig.~\ref{fig:variational_polaron}(d)). This then allows us to do the calculation in a much larger system with fixed $L_c$. Thus, it is convenient to plot our results for the energy as a function of $1/\sqrt{n}$. We see that our variational energy goes roughly as $1/\sqrt{n}$ and extrapolates to a value around $54\%$ of the single particle gap~\footnote{This value is sensitive to the data extrapolations: we only include data points that are linear in $1/\sqrt{n}$, which we have to exclude data for small $n$ (polaron limit) and large $n$ (finite size effect). Thus, the estimated extrapolated value is around $54\%$.}. This is very close to the value one obtained for an infinite Skyrmion in field theory, where $4\pi \rho \approx 54.50\%$ of the single particle gap for $d =10$ (this becomes exactly $50\%$ in the limit of unscreened Coulomb $d \rightarrow \infty$). This suggests that our ansatz captures both the limit of small and large spin polarons to very good accuracy.

\begin{figure}[t]
\centering
\includegraphics[width=\columnwidth]{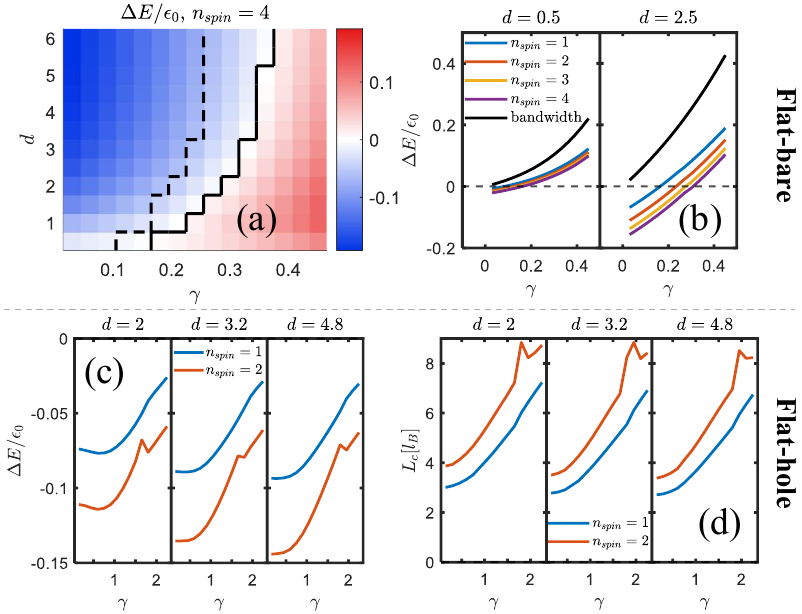}
\caption{Variational binding energy for polarons with different spin flips as functions of the non-uniformity $\gamma$ (see Eq.~\eqref{eq:non-uniformity}) and the screening length $d$ of the interaction for normal-ordered interaction with flat bare dispersion (top) and for flat hole dispersion (bottom). (a) Heatmap of the binding energy of the polarons with $4$ spin flips where the solid curve separates bound states (blue) and unbound states (red). The dashed curve indicates the boundary where the single-particle bandwidth exceeds the $\gamma = 0$ binding energy. (b) Line cuts of the binding eneregies at fixed screening length for different number of spin flips. We also add the single particle bandwidth. (c,d) Line cuts for the binding energies (c) and the optimal cut-off $L_c$ (d) at fixed screen lengths (d). We use a $17\times 17$ grid and $l_B^2=4\pi/3\sqrt{3}$ for all plots. }
\label{fig:spinpolaron_variational}
\end{figure}

\textit{Effect of quantum geometry on the polaron formation.---} Non-uniform quantum geometry influences spin polaron formation through two distinct mechanisms. First, it generates a quasihole dispersion whose bandwidth grows with the Berry curvature inhomogeneity and whose minimum sits at the curvature maximum. Second, in the limit where Berry curvature concentrates at a single point, the form factors reduce to those of a trivial band \cite{MottSemimetal}. Both effects disfavor binding. To quantify them, we tune the quantum geometry through a non-uniformity parameter $\gamma$ in the wavefunction~\eqref{eq:AC_wavefunction},
\begin{equation}
    Q(\br) = \gamma\sum_{\bG\in\text{first shell}}e^{i\bG\cdot\br},
    \label{eq:non-uniformity}
\end{equation}
where $\bG$ runs over the nonzero first-shell reciprocal lattice vectors. Through the relation between $Q(\br)$ and $\beta(\bk)$, increasing $\gamma$ concentrates the Berry curvature in momentum space.

We first consider a flat bare dispersion, $\epsilon(\bk)=0$, so that quantum geometry alone both generates the interaction-generated single hole dispersion (Fig.~\ref{fig:spinpolaron_variational}b) and influences the polaron wavefunction through the form factors. Figure~\ref{fig:spinpolaron_variational}a shows the binding energy at fixed spin-flip number $n_{\rm spin}$ as a function of the screening length $d$ and $s$: binding weakens monotonically with $\gamma$, and the phase boundary $\gamma_c(d)$ (black line) rises sharply at small $d$ before saturating. The $n_{\rm spin}=1,\dots,4$ line cuts in Fig.~\ref{fig:spinpolaron_variational}b confirm this trend and show, as in the uniform case, that bound states with more flipped spins are more robust against non-uniformity. We also show the bandwidth for interaction-generated dispersion for single holes which shows a very similar increase with increasing $\gamma$, suggesting dispersion is the predominant cause for the loss of binding. This is corroborated by identifying the boundary where the bandwidth of the interaction-generated dispersion equals the $\gamma=0$ binding energy indicated by the dashed line in Fig.~\ref{fig:spinpolaron_variational}a, which closely agrees with $\gamma_c(d)$. Both observations indicate that interaction-generated dispersion is the principal cause of unbinding: the polaron is heavy and benefits less from delocalization than a bare hole.

To isolate the geometric effects beyond single-particle dispersion, we now take $\epsilon(\bk)$ to exactly cancel the interaction-generated hole dispersion, yielding a flat hole band. Such cancellation is realized, for example, in TBG at hole-doped $\nu=1,2$, where the Hartree and Fock contributions compensate~\cite{Guinea2019, TBGV, PierceCDW, VafekBernevig}. As shown in Fig.~\ref{fig:spinpolaron_variational}c,d, binding still weakens and the bound-state size grows with $\gamma$, but the polaron now survives to much larger $\gamma$. This is naturally understood in the large-$\gamma$ limit, where the Berry curvature concentrates within a BZ patch of area $s^2 \ll 1$~\cite{MottSemimetal}. The associated geometric length $\sim 1/s$ sets a scale below which topology is invisible; such scale controls the bound-state size, with the binding energy vanishing as $s\to 0$ when the band becomes trivial.

\textit{Discussion: Implications for moir\'e Chern ferromagnets.}--- Our framework provides a route to analyze the energetics of non-trivial spin polarons in narrow Chern bands which emerge in moir\'e systems. One major advantage of our approach over the standard skyrmion field theory is that it allows us to encorporate momentum space structure into the ansatz, allowing to optimize for both quantum geometry variations and dispersion. While we derived our ansatz starting from the ideal band limit \cite{Ledwith2019, LandauLevelWang, LedwithVishwanathParker22}, our approach is ultimately variational which means that we can apply this ansatz for any band by taking $\beta(\bk)$ to be some arbitrary periodic function of $\bk$. This function can be either variationally optimized or taken to be defined by Eq.~\ref{deltaOmega} in terms of the Berry curvature. We note that since we work in a fixed $S_z$ sector, incorporating Zeeman in our framework is straightforward.

Our framework is applicable to any moir\'e platform with a flavor degenerate Chern $\pm 1$ band with approximate ${\rm SU}(2)$ flavor symmetry. This includes twisted bilayer graphene \cite{CaoIns, CaoSC, Yankowitz2019, Lu2019, Sharpe2019, Serlin2020}, twisted double bilayer \cite{KimTDBG, PabloTDBG, YankowitzTDBG} or mono-bilayer graphene \cite{YankowitzMonoBi1} and rhombohedral graphene \cite{YoungTrilayer2021, YoungRhombohedral2025, JuRhombohedralQAH} assuming an interaction-stabilized Chern band per valley \cite{DongAHC, SenthilAHC, YahuiAHC, TrithepAHC}. Chern Ferromagnets emerging in different platforms at various fillings will differ in three main aspects: (i) anisotropies which may generate effective Zeeman terms (e.g. intervalley Hund's coupling breaks independent spin symmetry per valley \cite{YahuiChernBands, Bultinck2020, ChatterjeeSkyrmion}), (ii) Berry curvature distribution (e.g. TBG is known to have a sharply concentrated Berry curvature at the $\Gamma$ point \cite{Ledwith2019,MottSemimetal}), and (iii) interaction-generated Hartree-Fock dispersion (e.g. the interaction-renormalized quasiparticle bands have much stronger dispersion when doping away from neutrality compared to doping towards neutrality in TBG \cite{Guinea2019, TBGV, PierceCDW, VafekBernevig}). Our results indicate that the stability of spin polarons depend on screening gate distance, Berry curvature inhomogeneity, effective Zeeman coupling, and quasiparticle dispersion. For instance, in TBG, we anticipate spin polarons to be more relevant on the heavy doping side ( towards neutrality) rather than the light doping side (away from neutrality) consistent with the findings of Refs.~\cite{KhalafBabySkyrmions, schindlerTrionsTwistedBilayer2022, KwanSkyrmion}. We leave a detailed analysis of specific material platforms for future works.

We close by noting that our construction for polaron operators and wavefunctions allows us to construct variational multipolaron states. For instance, we may consider the $m$-polaron variational states $|\psi \rangle = \int d^2 \bxi_1 \dots d^2 \bxi_m \psi(\bxi_1,\dots,\bxi_m) \hat P_{\bxi_1,n_1} \dots \hat P_{\bxi_m,n_m}\ket{\downarrow}$ where $\hat P_{\bxi,n} = \hat S_{\bxi}\,c_{\bxi,\downarrow}$ (cf.~Eq.~\ref{eq:polaronop}). This can be used to construct variational states for bound states of few spin polarons, e.g. Cooper pairs, or for phases of spin polarons like Wigner crystals or superconductors. This ansatz is expected to be a good description for phases in the limit of small doping when the polarons are energetically favored since it already accounts for the dressing effects from spin flips. We leave a more detailed study of such variational states to future works.

\emph{Acknowledgements}--- EK acknowledges discussions with Ashvin Vishwanath, Patrick Ledwith, and Daniel Parker. E.~K. is supported by NSF CAREER grant DMR award No. 2441781. The computations in this paper were run on the FASRC Cannon cluster supported by the FAS Division of Science Research Computing Group at Harvard University.

\begingroup
\renewcommand{\addcontentsline}[3]{}
\bibliographystyle{apsrev4-2} 
\bibliography{refs}
\endgroup

\clearpage
\renewcommand{\bibliography}[1]{}
\renewcommand{\bibliographystyle}[1]{}

\onecolumngrid
\setcounter{secnumdepth}{2}
\setcounter{page}{1}
\setcounter{equation}{0}
\setcounter{figure}{0}
\setcounter{table}{0}
\setcounter{section}{0}

\renewcommand{\theequation}{S\arabic{equation}}
\renewcommand{\thefigure}{S\arabic{figure}}
\renewcommand{\thetable}{S\arabic{table}}
\renewcommand{\thesection}{S\arabic{section}}
\makeatletter
\renewcommand\p@section{} 
\makeatother
\begin{center}
{\large\bf Supplemental Material for}\\
\vspace{0.5em}
{\it "Theory of microscopic spin polarons in Chern ferromagnets"}
\end{center}

\bigskip
\tableofcontents
\bigskip

\section*{Overview}
In this supplemental material, we first briefly review the construction of Landau levels on torus and the ideal bands. We then discuss the actions of the magnetic translation symmetry and the spin on the many-body wavefunctions. Later, we derive the main analytical results presented in the main text: the exact zero mode operator in the contact limit and the explicit wavefunction. In the end, we introduce the numerical method for evaluating the wavefunction overlaps.

\section{Review of Landau levels on the torus and Ideal bands}
\subsection{Setup and conventions}
\label{sec:setup}
 
We consider a two-dimensional system on a torus defined by lattice vectors $\bm{a}_1,\bm{a}_2$ with unit cell area $A_U = \mathrm{Im}(\bar a_1 a_2)$, where we use the complex notation $a_l = a_{l,x}+i\,a_{l,y}$. The reciprocal lattice vectors $\bm{b}_1,\bm{b}_2$ satisfy $\bm{a}_i\cdot\bm{b}_j = 2\pi\delta_{ij}$, and the Brillouin zone (BZ) has area $A_{\mathrm{BZ}} = (2\pi)^2/A_U$.
 
It is useful to define a dual effective magnetic field $\mathcal{B}$ in momentum space as
\begin{equation}
  \mathcal{B} \equiv \frac{A_U}{2\pi} = \frac{2\pi}{A_{\mathrm{BZ}}} = l_B^2,
  \label{eq:B_def}
\end{equation}
so that the magnetic length is $l_B = \sqrt{\mathcal{B}}$, while the real-space magnetic field is $B = 2\pi/A_U = \mathcal{B}^{-1}$. With one flux quantum per unit cell ($BA_U = 2\pi$), each Landau level contributes exactly one state per unit cell, consistent with $C=1$. As we will see, the dual field $\mathcal{B}$ is equal to the Berry curvature of the LLL and is the natural quantity appearing in the form factors and overlap integrals. In the following sections and in the main text, whenever we don't include the factor $\mathcal{B}$, we mean setting the magnetic length $l_B$ to be 1 and every other length scales are measured with respect to $l_B$, unless mentioned otherwise.
 
\subsection{LLL wavefunctions on the torus}
\label{sec:LLL}
 
\subsubsection{Weierstrass sigma function construction}
 
The LLL wavefunctions on the torus can be written in the compact form (up to a normalization factor)
\begin{equation}
  \varphi_{\bm{k}}(\bm{r})
  = e^{\frac{i}{2}z\bar k}\;
    \sigma\!\bigl(z + i\mathcal{B} k\,\big|\,a_1,a_2\bigr)\;
    e^{-\frac{1}{4\mathcal{B}}\,z\bar z},
  \label{eq:phi_k}
\end{equation}
where $z = x+iy$, $k = k_x+ik_y$, and $\sigma(z|a_1,a_2)$ is the Weierstrass sigma function associated to the lattice $L_c = \mathbb{Z}a_1+\mathbb{Z}a_2$. The sigma function is the unique entire, odd function with simple zeros at every lattice point and the quasi-periodicity property
\begin{equation}
  \sigma(z+R\,|\,a_1,a_2)
  = \eta_R\; e^{\frac{1}{2\mathcal{B}}\,\bar R\,(z+R/2)}\;\sigma(z\,|\,a_1,a_2),
  \label{eq:sigma_quasi}
\end{equation}
where $\eta_R = (-1)^{n+m+nm}$ for $R = n\,a_1+m\,a_2$. An explicit representation in terms of the Jacobi theta function is
\begin{equation}
  \sigma(z\,|\,a_1,a_2)
  = \frac{a_1}{\theta_1'(0)}\;
    \theta_1\!\Bigl(\frac{z}{a_1}\,\Big|\,\omega\Bigr)\;
    e^{\frac{\bar a_1}{4\mathcal{B} a_1}\,z^2},
  \qquad
  \omega = \frac{a_2}{a_1},
  \label{eq:sigma_theta}
\end{equation}
with $\theta_1(z|\omega) = \sum_{n=-\infty}^{\infty} e^{i\pi\omega(n+\frac{1}{2})^2 + 2\pi i(n+\frac{1}{2})(z+\frac{1}{2})}$.
 
The quasi-periodicity of $\sigma$ and the Gaussian factor $e^{-z\bar z/(4\mathcal{B})}$ conspire to give the correct magnetic boundary conditions:
\begin{equation}
  \hat T_\bR\,\varphi_{\bm{k}}(\bm{r})
  = \eta_R\; e^{i\bm{k}\cdot\bm{R}}\;\varphi_{\bm{k}}(\bm{r}),
  \label{eq:mag_BC}
\end{equation}
where $\hat T_\bR$ denotes the magnetic translation by lattice vector $R$. The magnetic translation is defined, in symmetric gauge, by
\begin{equation}
  \hat T_{\bm a}\,\psi(\bm r)
  \;=\; e^{-\frac{i}{2\mathcal B}\,\hat z\cdot(\bm a\times\bm r)}\,\psi(\bm r+\bm a)
  \;=\; e^{\,(a\bar z-\bar a z)/(4\mathcal B)}\,\psi(z+a,\bar z+\bar a),
  \label{eq:T_def}
\end{equation}
with $a = a_x+i a_y$. It commutes with the kinetic Hamiltonian for any $\bm a$, while different magnetic translations obey the projective algebra $\hat T_{\bm a}\hat T_{\bm b}=e^{\,i\hat z\cdot(\bm a\times\bm b)/(2\mathcal B)}\hat T_{\bm a+\bm b}$; the normalization $2\pi\mathcal B = A_U$ then guarantees that $\{\hat T_{\bm R}\}_{\bm R\in L_c}$ mutually commute, so that their simultaneous eigenstates can be labeled by a crystal momentum $\bm k$.
 
\subsubsection{Cell-periodic part and holomorphicity}
 
The cell-periodic (Bloch) part is
\begin{equation}
  u^{\mathrm{LLL}}_{\bm{k}}(\bm{r})
  = e^{-i\bm{k}\cdot\bm{r}}\,\varphi_{\bm{k}}(\bm{r})
  = e^{-\frac{i}{2}\bar z k}\;
    \sigma(z+i\mathcal{B} k\,|\,a_1,a_2)\;
    e^{-\frac{1}{4\mathcal{B}}\,z\bar z}.
  \label{eq:u_LLL}
\end{equation}
A crucial property is that $u^{\mathrm{LLL}}_{\bm{k}}(\bm{r})$ is an \textit{analytic function of $k$}---it depends on $k$ but not on $\bar k$, apart from the overall Gaussian envelope. This holomorphic structure is the hallmark of the LLL and is ultimately responsible for the saturation of the trace condition.
 
Under reciprocal lattice translations, the cell-periodic part satisfies
\begin{equation}
  u^{\mathrm{LLL}}_{\bm{k}+\bm{G}}(\bm{r})
  = \eta_{\bG}\; e^{-i\bm{G}\cdot\bm{r}}\;
    e^{\frac{\mathcal{B}}{2}\,\bar Gk}\;
    e^{\frac{\mathcal{B}}{4}\,G\bar G}\;
    u^{\mathrm{LLL}}_{\bm{k}}(\bm{r}).
  \label{eq:u_quasi_G}
\end{equation}
 
The LLL form factor is then given by:
\begin{equation}
  \lambda^{\mathrm{LLL}}_{\bm{q}}(\bm{k})
  = \frac{\langle u_{\bm{k}}|u_{\bm{k}+\bm{q}}\rangle}
         {\|u_{\bm{k}}\|\;\|u_{\bm{k}+\bm{q}}\|}
  = e^{-\frac{\mathcal{B}}{4}\,|q|^2\,-\,i\frac{\mathcal{B}}{2}\,\bm{q}\wedge\bm{k}},
  \label{eq:ff_LLL}
\end{equation}
which is the standard result: a Gaussian suppression at large $|\bm{q}|$ times a phase encoding the momentum-space magnetic translation algebra (the GMP algebra).

\subsection{Ideal Chern bands}
\label{sec:ideal}
 
\subsubsection{The trace condition}
 
For a general band with Bloch states $|u_{\bm{k}}\rangle$, the quantum geometric tensor $\mathcal{G}_{\alpha\beta}(\bm{k}) = \langle\partial_\alpha u_{\bm{k}}|\bigl(1-|u_{\bm{k}}\rangle\langle u_{\bm{k}}|\bigr)|\partial_\beta u_{\bm{k}}\rangle$ has a symmetric real part (the quantum metric $g_{\alpha\beta}$) and an antisymmetric imaginary part (the Berry curvature $\Omega$). These satisfy the general inequality
\begin{equation}
  \tr g(\bm{k}) \geq |\Omega(\bm{k})|.
  \label{eq:trace_ineq}
\end{equation}
An \textit{ideal Chern band} is one that saturates this bound at every $\bm{k}$:
\begin{equation}
  \tr g(\bm{k}) = |\Omega(\bm{k})|.
  \label{eq:trace_cond}
\end{equation}
For the LLL, this is trivially satisfied since $g = \frac{\mathcal{B}}{2}\,\mathbb{I}$ and $\Omega = \mathcal{B}$ are both constant. However, the condition can also be satisfied by bands with arbitrarily non-uniform Berry curvature.
 
The physical content of the trace condition is that the quantum geometric tensor is K\"ahler: the metric and curvature are locked together by holomorphicity, exactly as in the LLL. Equivalently, the wavefunctions retain a holomorphic structure in $k$, with all deviations from the LLL absorbed into smooth, position-dependent factors.
 
\subsubsection{Wavefunction parameterization}
 
The trace condition~\eqref{eq:trace_cond} is sufficiently constraining that, for $C=1$, it fixes the wavefunction to the form (unnormalized)
\begin{equation}
  u_{\bm{k}}(\bm{r})
  = u^{\mathrm{LLL}}_{\bm{k}}(\bm{r})\;\bm{\chi}(\bm{r})\;e^{\frac{1}{4}Q(\bm{r})},
  \label{eq:u_ideal}
\end{equation}
where $\bm{\chi}(\bm{r})$ is a normalized spinor ($\bm{\chi}^\dagger\bm{\chi}=1$, relevant when internal degrees of freedom are present) and $Q(\bm{r})$ is a real, periodic scalar function: $Q(\bm{r}+\bm{a}_l)=Q(\bm{r})$.
 
The physical interpretation is that an ideal $C=1$ band is equivalent to the LLL of a Dirac particle in a \textit{non-uniform} magnetic field
\begin{equation}
  B(\bm{r}) = B + \delta B(\bm{r}),
  \qquad
  \delta B(\bm{r}) = -\nabla^2 Q(\bm{r}),
  \label{eq:B_nonuniform}
\end{equation}
where $B = \mathcal{B}^{-1} = 2\pi/A_U$ is the uniform background field. The function $Q(\bm{r})$ is the \textit{K\"ahler potential} of the band. If $Q=0$, the band reduces to the uniform LLL. For moir\'e bands, $Q$ has the periodicity of the moir\'e lattice and can be expanded in reciprocal lattice harmonics:
\begin{equation}
  Q(\bm{r}) = \sum_{\bm{G}} Q_{\bm{G}}\, e^{i\bm{G}\cdot\bm{r}},
  \qquad
  \delta B(\bm{r}) = \sum_{\bm{G}} Q_{\bm{G}}\,|\bm{G}|^2\, e^{i\bm{G}\cdot\bm{r}}.
  \label{eq:Q_expansion}
\end{equation}
With a point group symmetry such as $C_6$, symmetry-distinct harmonics appear shell by shell, so the first few $Q_{\bm{G}}$ suffice to parameterize the band.
 
Now, we can define the non-negative periodic function
\begin{equation}
  K(\bm{r}) = e^{\frac{1}{2}Q(\bm{r})}
  = \sum_{\bm{G}} K_{\bm{G}}\, e^{i\bm{G}\cdot\bm{r}},
  \label{eq:K_def}
\end{equation}
where the Fourier coefficients $K_{\bm{G}}$ are determined by $Q_{\bm{G}}$ but are not equal to them. The dual function in momentum space is
\begin{equation}
  \mathcal{K}(\bm{k})
  = \sum_{\bm{G}} \eta_G\, K_{\bm{G}}\;
    e^{i\mathcal{B}\,\bm{G}\wedge\bm{k}}\;
    e^{-\frac{\mathcal{B}}{4}\,|G|^2}
  = \sum_{\bm{G}} \widetilde K_{\bm{G}}\;
    e^{i\mathcal{B}\,\bm{G}\wedge\bm{k}},
  \label{eq:Kcal_def}
\end{equation}
where $\widetilde K_{\bm{G}} = \eta_G\, K_{\bm{G}}\, e^{-\mathcal{B}|G|^2/4}$. The real-space and $k$-space functions are related by the integral transform
\begin{equation}
  \mathcal{K}(\bm{k})
  = e^{-\frac{\mathcal{B}}{2}|k|^2}\int d^2r\;|\sigma(z)|^2\;
    e^{-\frac{\mathcal{B}}{2}|r|^2}\;K(\bm{r})\;e^{i\bm{k}\cdot\bm{r}},
  \label{eq:K_transform}
\end{equation}
which is a Gaussian-weighted Fourier transform---essentially a Bargmann transform mapping position-space modulations to their momentum-space counterparts. It can also be understood as a twisted version of the Weierstrass transform \cite{YanAnyonDispersion}.
 
More generally, one defines a $\bm{q}$-dependent generalization
\begin{equation}
  \mathcal{K}_{\bm{q}}(\bm{k})
  = \sum_{\bm{G}} \eta_G\, K_{\bm{G}}\;
    e^{\frac{\mathcal{B}}{2}\,q\bar G}\;
    e^{i\mathcal{B}\,\bm{G}\wedge\bm{k}}\;
    e^{-\frac{\mathcal{B}}{4}\,|G|^2},
  \label{eq:Kq_def}
\end{equation}
which reduces to $\mathcal{K}(\bm{k})$ at $\bm{q}=\bm{0}$.
 
\subsubsection{Berry curvature and the reconstruction formula}
 
The norm of the Bloch state is $\|u_{\bm{k}}\|^2 \propto e^{\frac{\mathcal{B}}{2}|k|^2}\,\mathcal{K}(\bm{k})$, from which the Berry connection and curvature follow as
\begin{equation}
  \mathcal{A}(\bm{k}) = i\,\partial_{\bar k}\ln\|u_{\bm{k}}\|^2,
  \qquad
  \Omega(\bm{k}) = 2\,\partial_k\partial_{\bar k}\ln\|u_{\bm{k}}\|^2
  = \mathcal{B} + \delta\Omega(\bm{k}),
  \label{eq:berry}
\end{equation}
where $\delta\Omega(\bm{k}) = 2\,\partial_k\partial_{\bar k}\ln\mathcal{K}(\bm{k})$ averages to zero over the BZ. Expanding $\delta\Omega(\bm{k}) = \sum_{\bm{G}} \delta\Omega_{\bm{G}}\, e^{i\mathcal{B}\,\bm{G}\wedge\bm{k}}$, one can invert to obtain
\begin{equation}
  \mathcal{K}(\bm{k})
  = \exp\!\Biggl(\frac{2}{\mathcal{B}^2}\sum_{\bm{G}\neq\bm{0}}
    \frac{\delta\Omega_{\bm{G}}}{|\bm{G}|^2}\;
    e^{i\mathcal{B}\,\bm{G}\wedge\bm{k}}\Biggr).
  \label{eq:K_from_Omega}
\end{equation}
This is the practical bridge to numerics: given a band computed from a continuum or tight-binding model, one extracts $\Omega(\bm{k})$, Fourier-decomposes the fluctuation $\delta\Omega$, and constructs $\mathcal{K}(\bm{k})$---and hence all the form factors of the ideal band that best approximates the given band.
 
\subsection{Form factor of the ideal band}
\label{sec:factorization}
The overlap of ideal-band Bloch states can be written as
\begin{equation}
  \langle u_{\bm{k}}|u_{\bm{k}+\bm{q}}\rangle
  = e^{\frac{\mathcal{B}}{2}\,\bar k(k+q)}\;\mathcal{K}_{\bm{q}}(\bm{k}),
  \label{eq:overlap_ideal}
\end{equation}
from which the normalized form factor is
\begin{equation}
  \lambda_{\bm{q}}(\bm{k})
  = \lambda^{\mathrm{LLL}}_{\bm{q}}(\bm{k})\;\xi_{\bm{q}}(\bm{k}),
  \label{eq:ff_factor}
\end{equation}
where the LLL part is given by Eq.~\eqref{eq:ff_LLL} and the correction factor is
\begin{equation}
  \xi_{\bm{q}}(\bm{k})
  = \frac{\mathcal{K}_{\bm{q}}(\bm{k})}
         {\sqrt{\mathcal{K}(\bm{k})\;\mathcal{K}(\bm{k}+\bm{q})}}.
  \label{eq:xi_def}
\end{equation}

\subsection{Gauge choices}
\label{sec:gauge}
 
\subsubsection{Non-periodic gauge}
The natural gauge for the preceding analytic expressions is the non-periodic gauge, in which the creation operators satisfy
\begin{equation}
  c_{\bm{k}+\bm{G}} = e^{i\theta_{\bm{G}}(\bm{k})}\, c_{\bm{k}},
  \qquad
  \theta_{\bm{G}}(\bm{k}) = \frac{\mathcal{B}}{2}\,\bm{G}\wedge\bm{k}.
  \label{eq:nonper_gauge}
\end{equation}
The form factors transform correspondingly: $\lambda_{\bm{q}}(\bm{k}+\bm{G}) = e^{i[\theta_{\bm{G}}(\bm{k}+\bm{q})-\theta_{\bm{G}}(\bm{k})]}\,\lambda_{\bm{q}}(\bm{k})$. This gauge preserves the holomorphic structure and makes the sigma-function formulas natural.
 
\subsubsection{Periodic gauge}
For numerical calculations, one needs a \textit{periodic gauge} where $c^{\mathrm{per}}_{\bm{k}+\bm{G}} = c^{\mathrm{per}}_{\bm{k}}$. The transformation is
\begin{equation}
  c^{\mathrm{per}}_{\bm{k}}
  = \eta_{\{\bk\}}\;
    e^{i\frac{\mathcal{B}}{2}\,\{\bk\}\wedge[\bk]}\;c_{\bm{k}}\equiv\alpha_\bk c_{\bm{k}},
  \label{eq:per_gauge}
\end{equation}
where $[\bk]$ denotes the reduction of $\bm{k}$ to the first BZ and $\{\bm{k}\}=\bm{k}-[\bm{k}]$ is the reciprocal lattice part. One verifies $c^{\mathrm{per}}_{\bm{k}+\bm{G}} = c^{\mathrm{per}}_{\bm{k}}$ using the quasi-periodicity properties of $\eta$.
 
The form factors in the periodic gauge acquire additional phases. For $\bm{k}$ restricted to the first BZ,
\begin{equation}
  \lambda^{\mathrm{per}}_{\bm{q}}(\bm{k})
  = \eta_{\{\bk+\bq\}}\;
    e^{-i\frac{\mathcal{B}}{2}\,\{\bk+\bq\}\wedge[\bk+\bq]}\;
    \lambda_{\bm{q}}(\bm{k}).
  \label{eq:ff_per}
\end{equation}
These additional phases are straightforward to implement numerically and do not affect physical observables.

\subsection{Setting the non-uniformity}
\begin{figure}[t]
\centering
\includegraphics[width=0.9\columnwidth]{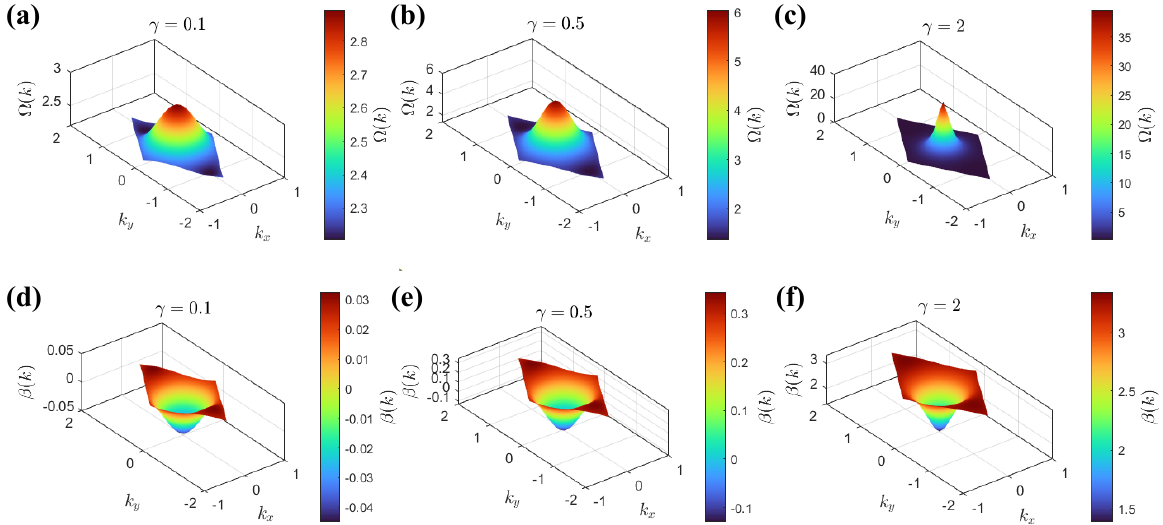}
\caption{The quantum geometry non-uniformity induced by $\gamma$: (a-c) the Berry curvature and (d-f) the corresponding $\beta(\bk)$.}
\label{fig:quantum_geometry}
\end{figure}

In the main text, we introduced a non-uniformity parameter $\gamma$ via:
\begin{equation}
    Q(\br) = \gamma\sum_{\bG\in\text{first shell}}e^{i\bG\cdot\br}
    \label{eq:non-uniformity}
\end{equation}
in the wavefunction~\eqref{eq:u_ideal} where $\bG$ is the first shell reciprocal lattice vector excluding $\boldsymbol{0}$. This non-uniformity also leads to a concentration of the Berry curvature through the relation between $Q(\br)$ and $\beta(\bk)$. Here $\beta(\bk) = \ln{\mathcal{K}(\bk)}$. We show the Berry curvature distributions and the $\beta(\bk)$ in Fig.~\ref{fig:quantum_geometry} for different $\gamma$'s.

\section{Action of magnetic translation symmetry}
Multiparticle excitations on top of the ferromagnetic state at half filling are obtained by creating some particles in the empty band and some holes in the filled band. Due to spin $\SU(2)$ symmetry, 
the number of particles and holes is separately conserved. As a result, the Hilbert space of $N_e$ electrons and $N_h$ holes is closed under the action of the Hamiltonian which allows the determination of few particle excitation spectrum exactly \cite{Lian2021, KhalafBabySkyrmions, schindlerTrionsTwistedBilayer2022}.

In this work, we study charge $\pm e$ excitations consisting of $n+1$ holes (electrons) and $n$ electrons (holes). For concreteness, we take the filled band to be $\downarrow$ and the empty to be $\uparrow$. We define the operators that create excitations relative to this reference state via
\begin{equation}
    d_{e,\bk} = c_{\uparrow,\bk}, \quad d_{h,\bk} = c^\dagger_{\downarrow,\bk}, \quad d_{\sigma, \bk + \bG} = \eta_\bG e^{-i \sigma \theta_\bG(\bk)} d_{\sigma, \bk}
\end{equation}
such that the ground state is annihilated by the operators $d_{\sigma,\bk}|\downarrow \rangle$ for all $\sigma = \pm \equiv e/h$, and $\bk$. 

A general state in the Hilbert space of $N_e$ $\uparrow$ electrons and $N_h$ $\downarrow$ holes is written as
\begin{equation}
    |\psi \rangle = \sum_{\{\bk_i\} \in {\rm BZ}} \psi(\bk_1, \dots,\bk_N) \prod_{i=1}^N d^\dagger_{ \sigma_i, \bk_i} | \downarrow \rangle
    \label{PsiKetN}
\end{equation}
where $N = N_e + N_h$ and $\sum_{i=1}^N \sigma_i = N_e - N_h$. $\psi(\bk_1,\dots,\bk_N)$ is antisymmetric under exchanging any two electrons or any two holes but is not constrained under exchanging an electron and a hole. In the following sections, we will sometimes label the electron momenta $\bk^e_l$, $l = 1,\dots, N_e$ and the hole momenta by $\bk^h_l$, $l = 1, \dots, N_h$. However, in this section, it will be convenient to use a common label for all momenta $\bk_l$, $l = 1,\dots,N$ and encode the electron/hole nature in the index $\sigma_l = \pm$. The requirement that the sum in (\ref{PsiKetN}) is independent of the choice of BZ imposes the boundary condition
\begin{equation}
    \psi(\bk_1, \dots, \bk_i + \bG,\dots, \bk_N) = \eta_\bG \psi(\bk_1, \dots, \bk_i,\dots, \bk_N) e^{-\frac{i}{2}\sigma_i \bG \wedge \bk_i}
    \label{PsiBC}
\end{equation}
The wavefunctions in the periodic gauge defined in~\ref{eq:per_gauge} can be obtained from $\psi$ via
\begin{equation}
    \psi^{\rm per}(\bk_1, \dots, \bk_N) = \psi(\bk_1, \dots, \bk_N) \prod_{i=1}^N \alpha_{\bk_i}^{-\sigma_i}
    \label{PsiPer}
\end{equation}

Now, let us discuss the action of magnetic translation symmetry on the many-body states in momentum space.
The many-body magnetic translation operator $\hat{T}_\bR$ acts on the excitation operators as
\begin{equation}
    \hat{T}_\bR\, d^\dagger_{\sigma,\bk}\, \hat{T}_\bR^{-1}
    = \eta_R\, e^{i\sigma\, \bk\cdot\bR}\, d^\dagger_{\sigma,\bk}.
    \label{TonD}
\end{equation}
For electrons ($\sigma = +$), this follows directly from the magnetic Bloch condition~\eqref{eq:mag_BC} applied to $d^\dagger_{e,\bk} = c^\dagger_{\uparrow,\bk}$. For holes ($\sigma = -$), conjugating $d^\dagger_{h,\bk} = c_{\downarrow,\bk}$ reverses the phase:
\begin{equation}
    \hat{T}_\bR\, c_{\downarrow,\bk}\, \hat{T}_\bR^{-1}
    = \overline{\eta_R\, e^{i\bk\cdot\bR}}\, c_{\downarrow,\bk}
    = \eta_R\, e^{-i\bk\cdot\bR}\, c_{\downarrow,\bk},
\end{equation}
where we used $\eta_R = \pm 1 \in \mathbb{R}$.

The reference state $|\!\downarrow\rangle = \prod_{\bk \in \mathrm{BZ}} c^\dagger_{\downarrow,\bk}\,|0\rangle$ transforms as
\begin{equation}
    \hat{T}_\bR\, |\!\downarrow\rangle
    = \eta_R^{N_s}\, e^{i\bK_0\cdot\bR}\, |\!\downarrow\rangle,
    \label{TonGS}
\end{equation}
where $N_s$ is the number of unit cells (equivalently, the number of $k$-points in the BZ) and $\bK_0 = \sum_{\bk \in \mathrm{BZ}} \bk$ is the total momentum of the filled band.

Combining~\eqref{TonD} and~\eqref{TonGS}, the translation acts on a general state~\eqref{PsiKetN} as
\begin{equation}
    \hat{T}_\bR\, |\psi\rangle
    = \eta_R^{N+N_s}\, e^{i\bK_0\cdot\bR}\,
      \sum_{\{\bk_i\} \in \mathrm{BZ}} \psi(\bk_1,\dots,\bk_N)\,
      e^{i\bigl(\sum_{j=1}^N \sigma_j \bk_j\bigr)\cdot\bR}\,
      \prod_{i=1}^N d^\dagger_{\sigma_i,\bk_i}\, |\!\downarrow\rangle.
    \label{TonPsi}
\end{equation}
Since $[H, \hat{T}_\bR] = 0$ for all lattice vectors $\bR$, the eigenstates can be labeled by a total crystal momentum $\bQ$ defined by the constraint
\begin{equation}
    \sum_{j=1}^N \sigma_j\, \bk_j
    = \sum_{l=1}^{N_e} \bk^e_l - \sum_{l=1}^{N_h} \bk^h_l
    = \bQ \quad (\mathrm{mod}\;\bG),
    \label{Ktot}
\end{equation}
on all configurations on which $\psi$ has support. A state with definite total momentum then satisfies
\begin{equation}
    \hat{T}_\bR\, |\psi_\bQ\rangle
    = \eta_R^{N+N_s}\, e^{i(\bK_0 + \bQ)\cdot\bR}\, |\psi_\bQ\rangle.
    \label{Teigenvalue}
\end{equation}
Since there is exactly one flux quantum per unit cell, the single-particle magnetic translations by lattice vectors commute, and consequently $\hat{T}_{\ba_1}$ and $\hat{T}_{\ba_2}$ commute as well. The total crystal momentum $\bQ$ therefore takes values on the discrete momentum grid within the BZ and provides a complete labeling of the translation symmetry sectors. The Hamiltonian is block-diagonal in $\bQ$.

\section{First quantized Hamiltonian in momentum space}
We consider a slightly different form of the (\ref{eq:Hproj})
\begin{equation}
\H=\sum_{\bk,\sigma} \tilde \epsilon(\bk) c_{\sigma,\bk}^\dagger c_{\sigma,\bk} + \frac{1}{2A}\sum_{\bq} V(\bq)\, \delta \rho_{-\bq} \delta \rho_{\bq},
\label{Hdelrho}
\end{equation}
where $\delta \rho_\bq = \rho_\bq - \xi_\bq$ and $\xi_\bq$ is a $c$-number representing the reference density per spin and $\tilde \epsilon(\bk)$ differs from $\epsilon(\bk)$ by single particle terms that compensate for normal ordering and density shift, which we provide explicitly below. First, $\xi_\bq$ is given by
\begin{equation}
    \xi_\bq = \delta_{[\bq],0} \sum_{\bk \in \BZ} \lambda^{\rm per}_{\bq}(\bk),
    \label{rq}
\end{equation}
where $[\bk]$ denotes the part of $\bk$ within the first BZ and $\{\bk\} = \bk - [\bk]$ denotes the reciprocal lattice part. This means that, as expected, $\xi_\bq$ vanishes unless $\bq$ is a reciprocal lattice vector. We emphasize that we need to use a periodic gauge such that $\lambda^{\rm per}_{\bq}(\bk + \bG) = \lambda^{\rm per}_{\bq}(\bk)$. The relation between $\tilde \epsilon(\bk)$ and $\epsilon(\bk)$ is obtained by noting that
\begin{equation}
    \frac{1}{2A}\sum_{\bq} V(\bq)\, \delta \rho_{-\bq} \delta \rho_{\bq} = \frac{1}{2A}\sum_{\bq} V(\bq)\, :\rho_{-\bq} \rho_{\bq}: + \sum_{\bk,\sigma} [\epsilon_F(\bk) - \epsilon_H(\bk)]
\end{equation}
where $\epsilon_F(\bk)$ and $\epsilon_H(\bk)$ are the Fock and Hartree energies given by
\begin{equation}
    \epsilon_F(\bk) = \frac{1}{2A} \sum_\bq V_\bq |\lambda_\bq(\bk)|^2, \qquad \epsilon_H(\bk) = \frac{1}{A} \sum_{\bG \in {\rm RL}} V_\bG \lambda_\bG^{{\rm per}}(\bk) \sum_{\bk' \in \BZ} \lambda_\bG^{{\rm per}}(\bk) 
\end{equation}
which implies that the choice
\begin{equation}
    \tilde \epsilon(\bk) = \epsilon(\bk) - \epsilon_F(\bk) + \epsilon_H(\bk)
\end{equation}
makes the Hamiltonian (\ref{Hdelrho}) equivalent to the one defined in Eq.~(\ref{eq:Hproj}) of the main text. To write the explicit form of the first quantized Hamiltonian in momentum space, we note
that the commutator of the $d$ operators with the density operator $\delta \rho_\bq$ is given by
\begin{gather}
    [\delta \rho_\bq, d^\dagger_{\sigma,\bk}] = \sigma \lambda^{(-\sigma )}_{-\sigma \bq}(\bk) d^\dagger_{\sigma,\bk - \sigma \bq}, \nonumber \\ \lambda^{(\pm)}_\bq(\bk) = |\lambda_\bq(\bk)| e^{\pm i \arg \lambda_\bq(\bk)}
    \label{rhod}
\end{gather}
At half-filling, it is straightforward to verify that any ferromagnet, e.g. $|\downarrow\rangle = \prod_\bk c_{\downarrow,\bk}^\dagger |0 \rangle$ (or any $\SU(2)$ rotated version of it), is a common eigenstate of all $\rho_\bq$ with eigenvalue $\xi_\bq$:
\begin{equation}
    \rho_\bq |\downarrow \rangle = \xi_\bq |\downarrow \rangle
\end{equation}
This means that $\delta \rho_\bq |\downarrow \rangle = 0$ for all $\bq$ and $\H |\downarrow \rangle = 0$. 

It is useful to define the operators
\begin{equation}
    T^i_\bq \psi(\bk_1, \dots, \bk_i,\dots, \bk_N) = \lambda^{(\sigma)}_\bq(\bk_i) \psi(\bk_1, \dots, \bk_i + \bq,\dots, \bk_N), 
\end{equation}
The action of the interaction part of the Hamiltonian (\ref{Hdelrho}) on the state $|\Psi \rangle$ can be obtained from the commutation relation (\ref{rhod}) 
giving
\begin{equation}
    \delta \rho_\bq |\psi \rangle = \sum_{\{\bk_i\}, l} \sigma_l [T^{l}_{\sigma_l \bq} \psi(\bk_1, \dots, \bk_N)] \prod_{i=1}^N d^\dagger_{ \sigma_i, \bk_i} | \downarrow \rangle 
\end{equation}
Thus, the action of the interacting part of the Hamiltonian is
\begin{equation}
    \H_{\rm int} |\psi \rangle = \sum_{\{\bk_i\}} [\hat V \psi(\bk_1, \dots, \bk_N)] \prod_{i=1}^N d^\dagger_{ \sigma_i, \bk_i} | 0 \rangle 
\end{equation}
where $\hat V$ is the first-quantized Hamiltonian given by
\begin{equation}
    \hat V = \frac{1}{2A} \sum_\bq V_\bq \hat O^\dagger_\bq \hat O_\bq ,\qquad
    \hat O_\bq = \sum_i \sigma_i T^i_{\sigma_i \bq}, 
    \label{VFirstQuantized}
\end{equation}

\begin{figure}[t]
\centering
\includegraphics[width=\columnwidth]{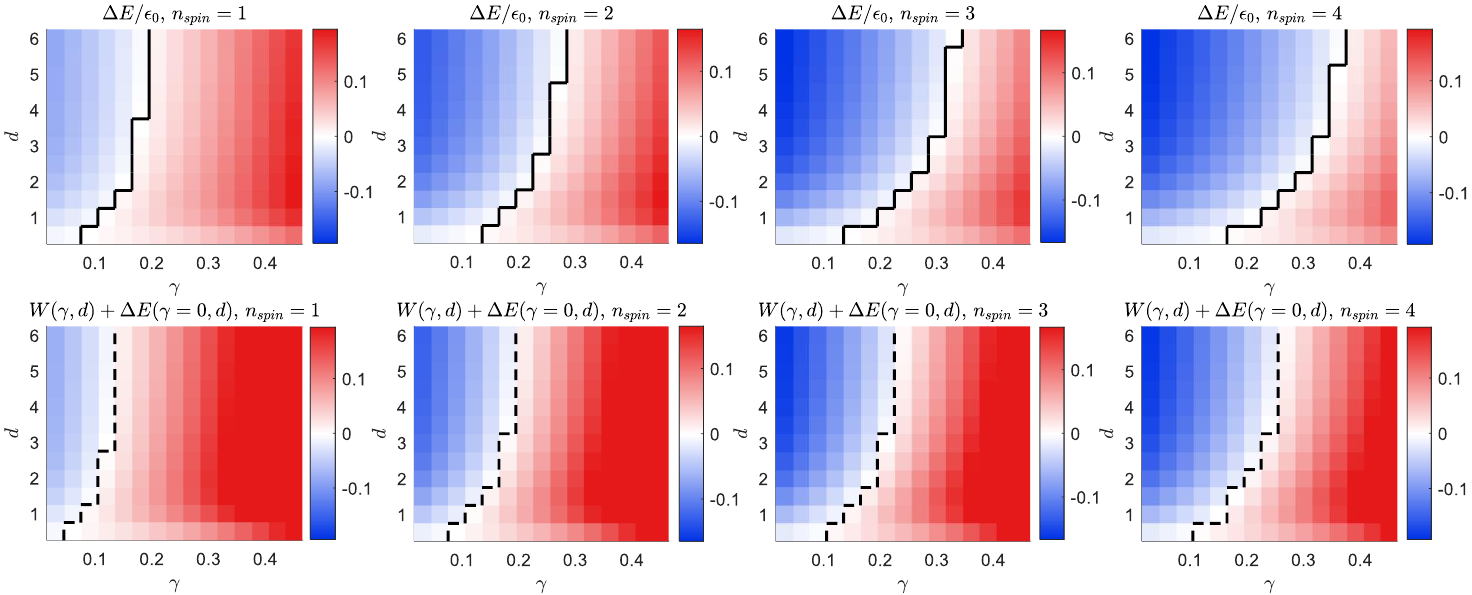}
\caption{The energy comparison for the normal-ordered interaction. Upper panels: the polaron binding energy $\Delta E(\gamma,d)$ for different numbers of spin flips. Lower panels: the shifted bandwidth $W(\gamma,d)+\Delta E(\gamma=0,d)$ for the same parameters.}
\label{fig:bandwidth_energy}
\end{figure}

It can be shown that for the normal-ordered density-density interaction, if $d_\text{screen}\to0$, the polaron should have the same energy as the hole. However, in this case, the hole has an interaction-induced dispersion (setting the bare dispersion to be flat): $2\epsilon_F-\epsilon_H$. To see how the dispersion affects the formation of the polaron, we compare the following two quantities: the binding energy $\Delta E(\gamma,d)$ and the shifted bandwidth $W(\gamma,d)+\Delta E(\gamma=0,d)$, where the later detects where the bandwidth will win over the strongest binding of the polarons. We show the comparison in Fig.~\ref{fig:bandwidth_energy}, where we see clearly that the band dispersion is the primary reason for the de-formation of the polarons.

\begin{figure}[t]
\centering
\includegraphics[width=0.55\columnwidth]{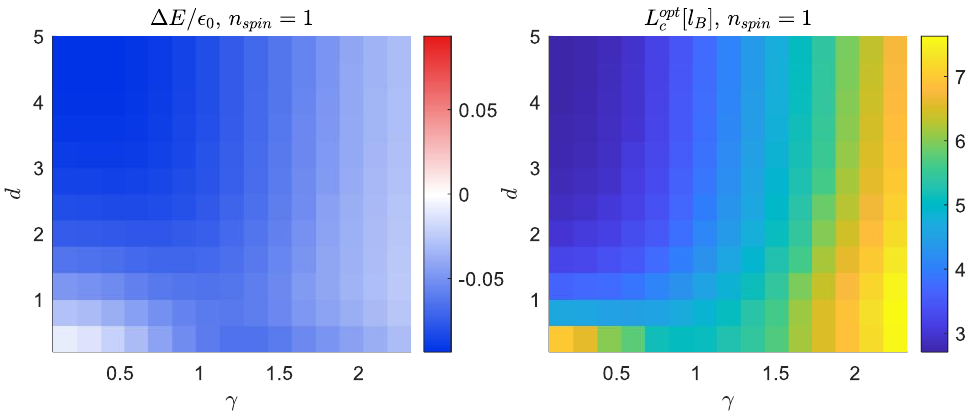}
\caption{Variational binding energy for polarons with different spin flips as functions of the non-uniformity $\gamma$ (see Eq.~\eqref{eq:non-uniformity}) and the screening length $d$ of the interaction. The bare dispersion is choosen such that the hole dispersion is flat. (left) Heatmap of the binding energy of the polarons with $1$ spin flips. (right) The optimal cut-off $L_c$ corresponding to data in the right panel.}
\label{fig:spinpolaron_variational_flat_hole}
\end{figure}

To isolate the effect of the dispersion, we now choose the bare dispersion to perfectly cancel this interaction-induced dispersion for the holes. The results for $n_\text{spin}=1$ are shown in Fig.~\ref{fig:spinpolaron_variational_flat_hole}.

\section{Action of spin}
\label{sec:spin}
The Hamiltonian~\eqref{Hdelrho} commutes with the global $\SU(2)$ generators $\bm{S} = (S^x,S^y,S^z)$, so the total spin quantum numbers $({\bm S}^2, S^z)$ provide additional labels for the eigenstates. In this section we work out the consequences for the excitation operators and many-body wavefunctions defined in the previous section.

\subsection{Global spin operators}
\label{sec:global_spin}
The global spin operators, projected to the two Chern bands, are
\begin{equation}
    S^z = \frac{1}{2}\sum_{\bk \in \mathrm{BZ}} \bigl(c^\dagger_{\uparrow,\bk}\, c_{\uparrow,\bk} - c^\dagger_{\downarrow,\bk}\, c_{\downarrow,\bk}\bigr),
    \qquad
    S^+ = \sum_{\bk \in \mathrm{BZ}} c^\dagger_{\uparrow,\bk}\, c_{\downarrow,\bk},
    \qquad
    S^- = (S^+)^\dagger.
    \label{eq:global_spin}
\end{equation}
These satisfy the standard $\SU(2)$ algebra $[S^z, S^\pm] = \pm S^\pm$, $[S^+,S^-] = 2S^z$, and commute with the Hamiltonian: $[H, S^a] = 0$ for all $a$. Note that the global spin operators ($\bq = 0$) do not involve form factors, since $\lambda_{\bq=0}(\bk) = \langle u_\bk | u_\bk \rangle = 1$.

In terms of the excitation operators $d_{e,\bk} = c_{\uparrow,\bk}$ and $d_{h,\bk} = c^\dagger_{\downarrow,\bk}$, the spin operators become
\begin{equation}
    S^+ = \sum_\bk d^\dagger_{e,\bk}\, d^\dagger_{h,\bk}, \qquad
    S^- = \sum_\bk d_{h,\bk}\, d_{e,\bk}.
    \label{eq:Spm_d}
\end{equation}
Thus $S^+$ creates an electron--hole pair at the same momentum (a $\bq = 0$ spin flip), while $S^-$ annihilates one.

\subsection{\texorpdfstring{$S^z$}{Sz} as a sector label}

The reference state $|\!\downarrow\rangle$ has all $N_s$ states in the $\downarrow$ band filled, giving
\begin{equation}
    S^z |\!\downarrow\rangle = -\frac{N_s}{2}\,|\!\downarrow\rangle.
\end{equation}
Each excitation operator $d^\dagger_{\sigma,\bk}$ raises $S^z$ by $\tfrac{1}{2}$, regardless of whether it creates an electron ($\sigma = +$) or a hole ($\sigma = -$):
\begin{equation}
    [S^z, d^\dagger_{\sigma,\bk}] = +\frac{1}{2}\, d^\dagger_{\sigma,\bk},
\end{equation}
since creating an $\uparrow$ electron contributes $+\tfrac{1}{2}$ directly, while creating a $\downarrow$ hole removes an electron of $S^z = -\tfrac{1}{2}$, also contributing $+\tfrac{1}{2}$.

A general state~\eqref{PsiKetN} with $N = N_e + N_h$ total excitations therefore has a definite $S^z$ eigenvalue
\begin{equation}
    S^z\,|\psi\rangle = \Bigl(-\frac{N_s}{2} + \frac{N}{2}\Bigr)|\psi\rangle
    = \Bigl(-\frac{N_s}{2} + \frac{N_e + N_h}{2}\Bigr)|\psi\rangle.
    \label{eq:Sz_eigenvalue}
\end{equation}
For charge-$e$ excitations with $N_h - N_e = 1$ and $n$ spin flips ($N_e = n$, $N_h = n+1$), the $S^z$ quantum number relative to the ground state is
\begin{equation}
    \Delta S^z = \frac{N_e + N_h}{2} = n + \frac{1}{2}.
\end{equation}

\subsection{Action of \texorpdfstring{$S^\pm$}{S±} on the many-body wavefunction}
\label{sec:Spm_action}

The operators $S^\pm$ connect adjacent sectors, mapping $(N_e, N_h) \to (N_e \pm 1, N_h \pm 1)$ while preserving the net charge $N_h - N_e$ and the total crystal momentum $\bQ$.

\subsubsection{\texorpdfstring{$S^+$}{S+}: raising}

Consider a state $|\psi\rangle$ in the $(N_e, N_h)$ sector with wavefunction $\psi(\bk^e_1,\dots,\bk^e_{N_e};\, \bk^h_1,\dots,\bk^h_{N_h})$, antisymmetric separately in the electron and hole momenta. From Eq.~\eqref{eq:Spm_d},
\begin{equation}
    S^+|\psi\rangle = \sum_\bp d^\dagger_{e,\bp}\, d^\dagger_{h,\bp}\, |\psi\rangle.
\end{equation}
To bring this into canonical form for the $(N_e{+}1,\, N_h{+}1)$ sector, we move $d^\dagger_{h,\bp}$ past the $N_e$ electron creation operators, incurring a sign $(-1)^{N_e}$. The resulting wavefunction is
\begin{equation}
    (S^+\psi)\bigl(\bk^e_1,\dots,\bk^e_{N_e+1};\,\bk^h_1,\dots,\bk^h_{N_h+1}\bigr)
    = (-1)^{N_e} \sum_{j=1}^{N_e+1}\sum_{l=1}^{N_h+1}
      (-1)^{j+l}\,\delta_{\bk^e_j,\,\bk^h_l}\;
      \psi\bigl(\widehat{\bk^e_j};\,\widehat{\bk^h_l}\bigr),
    \label{eq:Splus_psi}
\end{equation}
where $\widehat{\bk^e_j}$ (resp.\ $\widehat{\bk^h_l}$) denotes the list of electron (resp.\ hole) momenta with the $j$-th (resp.\ $l$-th) entry omitted. The Kronecker delta enforces that the new electron--hole pair shares the same momentum, as required by a $\bq = 0$ spin flip, and the signs arise from placing the $j$-th electron and $l$-th hole into the first position using the antisymmetry of $\psi$.

\subsubsection{\texorpdfstring{$S^-$}{S-}: lowering}

Similarly, $S^- = \sum_\bp d_{h,\bp}\, d_{e,\bp}$ maps $(N_e, N_h) \to (N_e{-}1,\, N_h{-}1)$. Its action on the wavefunction is
\begin{equation}
    (S^-\psi)\bigl(\bk^e_1,\dots,\bk^e_{N_e-1};\,\bk^h_1,\dots,\bk^h_{N_h-1}\bigr)
    = (-1)^{N_e-1}\sum_\bp
      \psi\bigl(\bp,\bk^e_1,\dots,\bk^e_{N_e-1};\,\bp,\bk^h_1,\dots,\bk^h_{N_h-1}\bigr),
    \label{eq:Sminus_psi}
\end{equation}
where the momentum $\bp$ is summed over the BZ. The operator $S^-$ traces over all ways to annihilate an electron--hole pair at the same momentum.

\subsubsection{Consistency with boundary conditions}

Since the global spin operators commute with the magnetic translations, $[\hat{T}_\bR, S^\pm] = 0$, the boundary condition~\eqref{PsiBC} is automatically preserved: if $\psi$ satisfies~\eqref{PsiBC}, then so do $S^+\psi$ and $S^-\psi$. The total crystal momentum $\bQ$ is unchanged by the action of $S^\pm$.

\subsection{Multiplet structure}

Since both $\bm{S}^2$ and $S^z$ commute with $H$, the eigenstates organize into $\SU(2)$ multiplets labeled by total spin $S$, with $2S+1$ degenerate states related by $S^\pm$. For a charge-$e$ excitation, a state in the $n$-spin-flip sector ($S^z = -N_s/2 + n + 1/2$) belongs to a multiplet with $S \geq n + 1/2$. The operators $S^\pm$ connect states in adjacent $n$-sectors within the same multiplet without changing the energy.

This has an important consequence: if $|\psi_n\rangle$ is an eigenstate of $H$ with energy $E$ in the $n$-sector, and $S^-|\psi_n\rangle \neq 0$, then $S^-|\psi_n\rangle$ is an eigenstate in the $(n{-}1)$-sector with the same energy $E$. Conversely, $S^+|\psi_n\rangle$ (if nonzero) is an eigenstate in the $(n{+}1)$-sector with energy $E$. States annihilated by $S^-$ are highest-weight states of their multiplet, and the dimension of the $n$-sector Hilbert space generally grows with $n$, introducing states that cannot be reached from lower sectors by $S^+$ alone.

\subsection{Magnon operators and the Goldstone mode}
\label{sec:magnon}

The spin-wave (magnon) operators at finite momentum $\bq$ involve the band form factors:
\begin{equation}
    S^+_\bq = \sum_\bk \lambda_\bq(\bk)\, c^\dagger_{\bk,\uparrow}\, c_{\bk+\bq,\downarrow}
    = \sum_\bk \lambda_\bq(\bk)\, d^\dagger_{e,\bk}\, d^\dagger_{h,\bk+\bq},
    \label{eq:magnon}
\end{equation}
where $\lambda_\bq(\bk) = \langle u_{\bk}|u_{\bk+\bq}\rangle$. The global spin raising operator is recovered at $\bq = 0$: $S^+_{\bq=0} = S^+$. For $\bq \neq 0$, the operator $S^+_\bq$ does not commute with $H$; it creates a magnon excitation carrying crystal momentum $\bq$.

Acting on the wavefunction, $S^+_\bq$ maps the $(N_e, N_h)$ sector to $(N_e{+}1, N_h{+}1)$ with a form-factor weight:
\begin{equation}
    (S^+_\bq\psi)\bigl(\bk^e_1,\dots,\bk^e_{N_e+1};\,\bk^h_1,\dots,\bk^h_{N_h+1}\bigr)
    = (-1)^{N_e}\sum_{j=1}^{N_e+1}\sum_{l=1}^{N_h+1}
      (-1)^{j+l}\,\lambda_\bq(\bk^e_j)\,\delta_{\bk^e_j,\,\bk^h_l+\bq}\;
      \psi\bigl(\widehat{\bk^e_j};\,\widehat{\bk^h_l}\bigr).
    \label{eq:magnon_psi}
\end{equation}
Compared to~\eqref{eq:Splus_psi}, the Kronecker delta now enforces $\bk^h_l = \bk^e_j + \bq$ (rather than $\bk^e_j = \bk^h_l$), and each term is weighted by $\lambda_\bq(\bk^e_j)$.

Since $\ket{\downarrow}$ spontaneously breaks the $\SU(2)$ spin symmetry, the magnon is a Goldstone mode with dispersion $\omega(\bq) \to 0$ as $\bq \to 0$. This leads to a key bound on the excitation spectrum. Let $|\psi_n\rangle$ be the lowest-energy charge-$e$ state in the $n$-spin-flip sector with energy $E_{\min}(n)$. The state $S^+_\bq |\psi_n\rangle$ lies in the $(n{+}1)$-sector with total momentum shifted by $\bq$, and its energy satisfies
\begin{equation}
    E_{\min}(n+1) \leq E_{\min}(n) + \omega(\bq).
    \label{eq:Goldstone_bound}
\end{equation}
Taking $\bq \to 0$, we conclude
\begin{equation}
    E_{\min}(n+1) \leq E_{\min}(n).
    \label{eq:Emin_monotone}
\end{equation}
The minimum excitation energy is therefore non-increasing with the number of spin flips $n$.

It is important to distinguish two limits. At exactly $\bq = 0$, the operator $S^+_{\bq=0} = S^+$ is the global spin raising operator, which simply rotates the state within its $\SU(2)$ multiplet: $S^z$ increases by 1 but $\bm{S}^2$ is unchanged, and the energy is exactly $E_{\min}(n)$. This does not produce a genuinely new state in the $(n{+}1)$-sector with independent quantum numbers. For $\bq$ infinitesimal but nonzero, however, $S^+_\bq$ creates a magnon with nontrivial spatial structure, yielding a state with \textit{different} $\bm{S}^2$ (since $S^+_{\bq\neq0}$ doesn't commute with $\bm{S}^2$) and energy $E_{\min}(n) + \omega(\bq) \to E_{\min}(n)$. It is this latter limit that establishes~\eqref{eq:Emin_monotone}.

\section{Derivation of the zero mode operator}
We start by considering the most general form for a charge $e$, spin polaron consisting of two holes and one electron given by the wavefunction
\begin{equation}
    |\psi\rangle = \hat O |\downarrow \rangle = \sum_{\bk^e, \bk^h_1, \bk^h_2} \psi(\bk^e, \bk^h_1, \bk^h_2) c_{\bk^e, \uparrow}^\dagger c_{\bk^h_1,\downarrow} c_{\bk^h_2, \downarrow} |\downarrow \rangle
\end{equation}
where $\psi(\bk^e, \bk^h_1, \bk^h_2)$ is antisymmetric under exchanging the hole momenta $\bk^h_1$ and $\bk^h_2$. A sufficient condition for $|\psi \rangle$ to be a zero mode for $\H_{\rm NO}$ is 
\begin{gather}
    \label{OMCommutator}
    [\hat O, \M(\br)] = 0 = \sum_{\bk,\bk^h_1,\bk^h_2} R_{\bk,\bk^h_1,\bk^h_2}(\br) c_{\bk, \downarrow} c_{\bk^h_1,\downarrow} c_{\bk^h_2, \downarrow} \\
    \label{R3R2}
   R_{\bk,\bk^h_1,\bk^h_2}(\br) = \phi_{\bk}(\br) R_{\bk^h_1,\bk^h_2}(\br),\\ R_{\bk^h_1,\bk^h_2}(\br) = \sum_\bk \phi_{\bk}(\br) \psi(\bk,\bk^h_1,\bk^h_2)
\end{gather}
The last equation implies that $R_{\bk^h_1,\bk^h_2}$ belongs to the LLL, $P^{\rm LLL} R_{\bk^h_1,\bk^h_2} = R_{\bk^h_1,\bk^h_2}$. Furthermore, using the orthonormality of the Bloch states, we can solve it to get
\begin{equation}
    \psi(\bk^e, \bk^h_1, \bk^h_2) = \int d^2 \br \phi^*_{\bk^e}(\br) R_{\bk^h_1, \bk^h_2}(\br)
    \label{PsiInt}
\end{equation}
Eq.~\ref{OMCommutator} is fulfilled if and only $R_{\bk,\bk^h_1,\bk^h_2}(\br)$ vanishes upon antisymmetrizing relative to its three variables which, together with (\ref{R3R2}), implies
\begin{equation}
    R_{\bk^h_1,\bk^h_2}(\br) = \A \phi_{\bk^h_1}(\br) f_{\bk^h_2}(\br) 
    \label{Rk}
\end{equation}
where $\A$ is the antisymmetrization operator relative to the hole momenta $\bk^h_1$ and $\bk^h_2$.
The requirement that $R_{\bk_2, \bk_3}$ is in the LLL places a restriction on $f_\bk(\br)$. One obvious possibility is to choose $f_\bk(\br)$ to be holomorphic in $r$ such that $f_\bk(r) \phi_{\bk_2}(\br)$ lies in the LLL. To obtain a finite result in (\ref{PsiInt}), we require the integrand to not grow at $\infty$ which implies that $f_{\bk}(r)$ is a constant in $\br$ leading to
\begin{equation}
    \psi(\bk^e, \bk^h_1, \bk^h_2) = \A f_{\bk^h_2} \int d^2 \br \phi^*_{\bk^e}(\br) \phi_{\bk^h_1}(\br) = \A f_{\bk^h_2} \delta(\bk^e - \bk^h_1)
\end{equation}
Here, we have assumed $\bk^h_{1,2}$ are in the first BZ. It is easy to extend this to a wavefunction satisfying the boundary condition by summing over reciprocal lattice shifts. The wavefunction above represents a single hole attached to a Goldstone mode as can be verified by computing its $S^2$ eigenvalue as described in Sec.~\ref{sec:spin}. This means this is \textit{not} the non-trivial polaron wavefunction we are looking for.

Due to the antisymmetrization operator, there is another possible choice which ensures $R_{\bk^h_1,\bk^h_2}$ lies in the LLL. Consider the function $\psi_\xi(z) = \frac{f(z)}{z - \xi} e^{-\frac{1}{4}|z|^2}$ for $f(z)$ is holomorphic. The action of the LLL projection on this function is (see Sec.~\ref{App:LLLMomentum} for details)
\begin{equation}
    [P^{\rm LLL} \psi_\xi](z) = \frac{e^{-\frac{1}{4}|z|^2}}{z - \xi}[f(z) - f(\xi)] 
\end{equation}
The second term above means that $\psi_\xi(z)$ is not a LLL wavefunction. However, choosing $f_\bk(\br) = \frac{1}{r - \xi} \phi_\bk({\vec \xi})$, we see that $R_{\bk_2, \bk_3}$ is a LLL wavefunction due to the cancellation of the second term by antisymmetry. This leads to the polaron wavefunction in the main text.

\section{Derivation of explicit wavefunction expressions}
\label{App:MomentumWavefunctionDerivation}

In this appendix, we will derive the expressions for the polaron wavefunction (\ref{eq:psi_general}) and (\ref{PsiNGeminal}). For simplicity, we will begin with the single spin flip case. Generalizations to the multi-spin flip case is straightforward. We start with the expression for the polaron wavefunction obtained by substituting $\psi_\xi(z) = \frac{f(z)}{z - \xi} e^{-\frac{1}{4}|z|^2}$ for $f(z)$ in Eq.~\ref{Rk} and substituting that in Eq.~\ref{PsiInt}, leading to 
\begin{equation}
    \psi_\xi(\bk_1, \bk_2, \bk_3) = \A \phi_{\bk_3}({\vec \xi}) \int d^2 \br \frac{1}{r - \xi} \phi^*_{\bk_1}(\br) \phi_{\bk_2}(\br)
\end{equation}
This expression can be further simplified by writing $\br = \bR + \bu$ where $\bu$ is inside the unit cell and $\bR$ is a lattice vector. We now use the property $\phi_\bk(\br + \bR) = \eta_\bR e^{i \bk \cdot \bR} e^{\frac{i}{2} \bR \wedge \br} \psi_\bk(\br)$ to obtain
\begin{equation}
\psi_\xi(\bk_1, \bk_2, \bk_3) = \A \phi_{\bk_3}({\vec \xi}) \int d^2 \bu \phi^*_{\bk_1}(\bu) \phi_{\bk_2}(\bu) \gamma(u - \xi, \bk_2 -  \bk_1), \qquad \gamma(u,\bk) = \sum_\bR \frac{1}{u + R}  e^{i \bk \cdot \bR} 
\label{Psigamma}
\end{equation}
Instead of directly evaluating the summation in $\gamma(u,\bk)$, we will use its periodic and analytic structure to deduce its form by invoking Liouville's theorem. Let us start by defining $\tilde \gamma(u,\bk) = \gamma(u,\bk) e^{\frac{i}{2}u \bar k}$. We now see that for $u$ inside the unit cell and $\bk$ inside the first BZ, $\tilde \gamma(u,\bk)$ satisfies
\begin{equation}
    -2i \partial_{\bar k} \tilde \gamma(u,\bk) = \sum_\bR e^{i \bk \cdot \bR} \propto \delta(\bk), \qquad \partial_{\bar u} \tilde \gamma(u,\bk) \propto \delta(u)
    \label{dgamma}
\end{equation}
In addition, $\tilde \gamma(u,\bk)$ satisfies the boundary conditions 
\begin{equation}
    \tilde \gamma(u + R, \bk) = e^{-\frac{i}{2} \bar R k} \tilde \gamma(u, \bk) \qquad \tilde \gamma(u, \bk + \bG) = e^{\frac{i}{2} u \bar G} \tilde \gamma(u, \bk)
    \label{gammaBC}
\end{equation}
for lattice vector $\bR$ and reciprocal lattice vector $\bG$. Eqs.~\ref{dgamma} can be solved by writing $\tilde \gamma(u,\bk) \propto \frac{\sigma(k + i u)}{\sigma(k) \sigma(i u)} F(u,k)$ where $F$ is holomorphic in both its arguments. The boundary conditions (\ref{gammaBC}) implies $F(u+R,k) = F(u, k+G) = F(u,k)$, thus by Liuoville's theorem, $F$ is a constant. This yields
\begin{equation}
    \gamma(u,\bk) \propto \frac{e^{-\frac{i}{2}u \bar k} \sigma(k + i u)}{\sigma(k) \sigma(i u)}
\end{equation}

Substituting in (\ref{Psigamma}), we get
\begin{align}
\psi_\xi(\bk_1, \bk_2, \bk_3) &\propto  \A \phi_{\bk_3}({\vec \xi}) \int d^2 \bu \phi^*_{\bk_1}(\bu) \phi_{\bk_2}(\bu) \frac{\sigma(k_2 - k_1 + i (u - \xi)) e^{-\frac{i}{2}(u - \xi)(\bar k_2 - \bar k_1)}}{\sigma(k_2 - k_1) \sigma(i (u - \xi))} \nonumber \\
&\propto e^{\frac{i}{2} (\bar k_3 + \bar k_2 - \bar k_1) \xi} e^{-\frac{1}{4}(\bk_1^2 + \bk_2^2 + \bk_3^2)}\\
&\qquad\times\A \left\{ \sigma(k_3 - i \xi)   \int d^2 \bu \, e^{-\frac{1}{2}\bu^2 + \frac{i}{2}(u \bar k_1 - \bar u k_1)} \bar \sigma(k_1 - i u) \sigma(k_2 - i u) 
\frac{\sigma(k_2 - k_1 + i (u - \xi)) }{\sigma(k_2 - k_1) \sigma(i (u - \xi))} \right\} \nonumber \\
&= e^{\frac{i}{2} (\bar k_3 + \bar k_2 - \bar k_1) \xi} e^{-\frac{1}{4}(\bk_1^2 + \bk_2^2 + \bk_3^2)}    \int d^2 \bu \, e^{-\frac{1}{2}\bu^2 + \frac{i}{2}(u \bar k_1 - \bar u k_1)} | \sigma(k_1 - i u)|^2 F_{u,\xi}(\bk_1, \bk_2, \bk_3) \nonumber \\
&= e^{\frac{i}{2} (\bar k_3 + \bar k_2 - \bar k_1) \xi} e^{\frac{1}{4}(\bk_1^2 - \bk_2^2 - \bk_3^2)} \int d^2 \bu \, e^{-\frac{1}{2}|u + i k_1|^2} | \sigma(k_1 - i u)|^2 F_{u,\xi}(\bk_1, \bk_2, \bk_3) 
\label{PsiFxi}
\end{align}
where 
\begin{equation}
    F_{u,\xi}(\bk_1, \bk_2, \bk_3) = \A \left\{\frac{\sigma(k_2 - k_1 + i u - i\xi) \sigma(k_3 - i \xi) \sigma(k_2 - i u)}{\sigma(k_2 - k_1) \sigma(i u - i \xi) \sigma(k_1 - i u)} \right\}
    \label{Fuxi}
\end{equation}
We can easily show from the behavior of the $\sigma$ function under reciprocal lattice translations that $F_{u,\xi}$ is a periodic function in $u$. Second, notice that the residues of the poles at $u = \xi$ and $u = -i k_1$ vanish upon antisymmetrization. Thus, by Liouville's theorem, this function is a constant in $u$: $F_{u,\xi}(k_1, k_2, k_3) = F_\xi(k_1, k_2, k_3)$. 
Thus, we can simplify (\ref{PsiFxi}) as
\begin{align}
    \psi_\xi(\bk_1, \bk_2, \bk_3) &\propto e^{\frac{i}{2} (\bar k_3 + \bar k_2 - \bar k_1) \xi} e^{\frac{1}{4}(\bk_1^2 - \bk_2^2 - \bk_3^2)} F_{\xi}(\bk_1, \bk_2, \bk_3) \int d^2 \bu \, e^{-\frac{1}{2}|u + i k_1|^2} | \sigma(k_1 - i u)|^2 \nonumber \\
    &\propto e^{\frac{i}{2} (\bar k_3 + \bar k_2 - \bar k_1) \xi} e^{\frac{1}{4}(\bk_1^2 - \bk_2^2 - \bk_3^2)} F_{\xi}(\bk_1, \bk_2, \bk_3) 
    \label{PsiFFinal}
\end{align}
Here we used the fact that the integral over $\bu$ only depends on the combination $u+i k_1$ which means we can get rid of the dependence on $k_1$ by the variable shift $u \rightarrow u - i k_1$.
We can evaluate $F_\xi(k_1, k_2, k_3)$ from $F_{u,\xi}(k_1, k_2, k_3)$ in (\ref{Fuxi}) by choosing any fixed value of $u$. We choose as $i u = k_3$ leading to
\begin{align}
    F_\xi(k_1, k_2, k_3) &= F_{u=-i k_3, \xi}(k_1, k_2, k_3) \nonumber \\
    &= \left\{\frac{\sigma(k_2 - k_1 + i u - i\xi) \sigma(k_3 - i \xi) \sigma(k_2 - i u)}{\sigma(k_2 - k_1) \sigma(i u - i \xi) \sigma(k_1 - i u)} - \frac{\sigma(k_3 - k_1 + i u - i\xi) \sigma(k_2 - i \xi) \sigma(k_3 - i u)}{\sigma(k_3 - k_1) \sigma(i u - i \xi) \sigma(k_1 - i u)}\right\}\Big|_{u = -i k_3}
    \nonumber \\&= \frac{\sigma(k_1 - k_2 - k_3 + i\xi) \sigma(k_2 - k_3)}{\sigma(k_2 - k_1)  \sigma(k_3 - k_1)} 
\end{align}
Substituting in (\ref{PsiFFinal}) gives Eq.~\ref{eq:psi_general} of the main text.

We now want to derive Eq.~\ref{PsiNGeminal} of the main text. From Eq.~\ref{PsiFxi}, we see that the wavefunction has the form (\ref{PsiNGeminal} with $g_\bxi(\bk_3) = \phi_{\bk_3}(\bxi)$ and $f_\bxi(\bk_1, \bk_2)$ given by
\begin{equation}
    f_\bxi(\bk_1, \bk_2) = e^{\frac{i}{2} ( \bar k_2 - \bar k_1) \xi} e^{\frac{1}{4}(\bk_1^2 - \bk_2^2)} \int d^2 \bu \, e^{-\frac{1}{2}|u + i k_1|^2} | \sigma(k_1 - i u)|^2 M_{u,\xi}(\bk_1, \bk_2) 
    \label{fxiIntegral}
\end{equation}
where $M_{u,\xi}(\bk_1, \bk_2)$ is
\begin{equation}
    M_{u,\xi}(\bk_1, \bk_2) = \frac{\sigma(k_2 - k_1 + i u - i\xi)  \sigma(k_2 - i u)}{\sigma(k_2 - k_1) \sigma(i u - i \xi) \sigma(k_1 - i u)}
\end{equation}
It is straightforward to verify that this function is periodic in $u$. This means that it is an elliptic function of $u$ poles at $u = \xi$ and $u = -i k_2$. Below, we will derive a simple form of this function. Let us introduce the Fourier transform in $u$:
\begin{equation}
    M_{\bG,\xi}(\bk_1, \bk_2) = \int_{\rm UC} \frac{d^2 u}{2\pi} e^{i \bG \cdot \bu} M_{u,\xi}(\bk_1, \bk_2)
\end{equation}
For $\bG \neq 0$, we can evaluate this as
\begin{align}
    M_{\bG,\xi}(\bk_1, \bk_2) &= \int_{\rm UC} \frac{d^2 \bu}{2\pi} \frac{2}{i G} [\partial_{\bar u} e^{i \bG \cdot \bu}] M_{u,\xi}(\bk_1, \bk_2) = -\frac{1}{i \pi G} \int_{\rm UC} d^2 \bu \,  e^{i \bG \cdot \bu} \partial_{\bar u} M_{u,\xi}(\bk_1, \bk_2) \nonumber \\
    &= \frac{1}{G \sigma'(0)} \frac{\sigma(k_2 - i \xi)}{\sigma(k_1 - i \xi)} [e^{i \bG \cdot \bxi} - e^{i \bG \wedge \bk_1}]
\end{align}
Substituting in (\ref{fxiIntegral}), we notice that the contribution coming from $\bG \neq 0$ gives a term with the form $g_\bxi(\bk_2) h_{\bxi}(\bk_1)$ which does not contribute to the wavefunction vanishes upon antisymmetrization and thus do not contribution to the wavefucntion. Thus, we can drop that term and only consider the $\bG = 0$ contribution. This can be evaluated by noting that $M$ vanishes when $k_2 = i u$ to write
\begin{equation}
    0 = M_{\bG=0,\xi}(\bk_1, \bk_2) + \sum_{\bG \neq 0} e^{-i \bG \wedge \bk_2} M_{\bG,\xi}(\bk_1, \bk_2)
\end{equation}
which gives
\begin{align}
    M_{\bG=0,\xi}(\bk_1, \bk_2) &= -\sum_{\bG \neq 0} e^{i \bG \wedge \bk_2} M_{\bG,\xi}(\bk_1, \bk_2) \propto \frac{\sigma(k_2 - i \xi)}{\sigma(k_1 - i \xi)} \sum_{\bG \neq 0} \frac{e^{i \bG \cdot (\bxi - \wedge \bk_2)} - e^{i \bG \wedge (\bk_1 - \bk_2)}}{G} \nonumber \\
    & \propto \frac{\sigma(k_2 - i \xi)}{\sigma(k_1 - i \xi)} \sum_{\bR \neq 0} \frac{e^{-i \bR \cdot (\wedge \bxi + \bk_2)} - e^{i \bR \cdot (\bk_1 - \bk_2)}}{R} \nonumber \\
    & \propto \frac{\sigma(k_2 - i \xi)}{\sigma(k_1 - i \xi)} \lim_{z \rightarrow 0} [\gamma(z,-\wedge \bxi - \bk_2) - \gamma(z,\bk_1 - \bk_2)] \nonumber \\
    &\propto \frac{\sigma(k_2 - i \xi)}{\sigma(k_1 - i \xi)} [\zeta(k_2 - i\xi) - \zeta(k_2 - k_1)]
\end{align}
where we introduced the periodic Weierstrass $\zeta$ function
\begin{equation}
    \zeta(k) = \frac{\sigma'(k)}{\sigma(k)} - \frac{1}{2} \bar k
\end{equation}
Substituting in~\eqref{fxiIntegral} leads to Eq.~\eqref{PsiNGeminal} in the main text.

\section{Action of LLL projector}
\label{App:LLLMomentum}
Consider the normalized LLL Bloch states given by
\begin{equation}
    \phi_\bk(\br) = C(\bk) e^{-\frac{1}{4} \bk^2} \varphi_\bk(\br) 
\end{equation}
where $\varphi_\bk(\br)$ are the unnormalized Bloch states (whose periodic part is holomorphic in $k$) introduced in Eq.~\ref{eq:phi_k}. Here, we have introduced the $\bk$-dependent normalization $C(\bk)$ pulled out a gaussian factor in $\bk$ for latter convenience.
The overlap of two Bloch states can be evaluated as
\begin{align}
    \langle \phi_{\bk_1}|\phi_{\bk_2} \rangle &=  \frac{|C(\bk)|^2}{A} \int d^2 \br \bar \phi_{\bk_1}(\br) \phi_{\bk_2}(\br) = \frac{|C(\bk)|^2}{A} \int d^2 \br e^{-\frac{1}{2} r \bar r - \frac{1}{4}(k_1 \bar k_1 + k_2 \bar k_2)} e^{\frac{i}{2}(r \bar k_2 - \bar r k_1)} \bar \sigma(r + i k_1) \bar \sigma(r + i k_2) \nonumber \\
    &= \frac{|C(\bk)|^2}{A} e^{-\frac{1}{2} k_2 \bar k_1 - \frac{1}{4}(k_1 \bar k_1 + k_2 \bar k_2)} \int d^2 \br |\sigma(r)|^2 e^{-\frac{1}{2} r \bar r} e^{i \br \cdot (\bk_2 - \bk_1)}
\end{align}
where $A$ is the total area $A = N A_{\rm UC} = 2\pi N$. In going from the first to the second line, we have performed the variable shift $r \mapsto r - i k_2$, $\bar r \mapsto \bar r + i \bar k_1$. The function $\sigma(r) e^{-\frac{1}{4} r \bar r}$ changes by a phase under translation which implies that $|\sigma(r)|^2 e^{-\frac{1}{2} r \bar r}$ is periodic. We can now split $\br$ into a part within the unit cell $\bu$ and a lattice vector $\bR$ leading to
\begin{gather}
    \frac{1}{A} \int d^2 \br |\sigma(r)|^2 e^{-\frac{1}{2} r \bar r} e^{i \br \cdot (\bk_2 - \bk_1)} = \frac{1}{2\pi N} \int d^2 \bu |\sigma(u)|^2 e^{-\frac{1}{2} u \bar u} e^{i \bu \cdot (\bk_2 - \bk_1)} \sum_\bR e^{i \bR \cdot (\bk_1 - \bk_2)} = \delta_{\bk_1, \bk_2} F_0 \\
    F_0 = \frac{1}{ 2\pi} \int d^2 \bu |\sigma(u)|^2 e^{-\frac{1}{2} u \bar u} 
\end{gather}
Thus, by choosing $|C(\bk)|=F_0^{-1/2}$, we get the orthogonality relation we expect
\begin{equation}
    \langle \phi_{\bk_1}|\phi_{\bk_2} \rangle = \frac{1}{A} \int d^2 \br \bar \phi_{\bk_1}(\br) \phi_{\bk_2}(\br) =  \delta_{\bk_1, \bk_2}
\end{equation}
The band projector is given by
\begin{equation}
    P(\br_1, \br_2) := \frac{1}{2\pi} \int_{\rm BZ} d^2 \bk \phi_\bk(\br_1) \bar \phi_\bk(\br_2) = \frac{F_0^{-1}}{2\pi} \int_{\rm BZ}  d^2 \bk e^{-\frac{1}{2} k \bar k - \frac{1}{4} (r_1 \bar r_1 + r_2 \bar r_2)} e^{\frac{i}{2}(r_1 \bar k - \bar r_2 k)} \sigma(ik + r_1) \bar \sigma(ik + r_2)
\end{equation}
We now shift the integration variables $k \mapsto k + i r_1$ and $\bar k \mapsto \bar k - i \bar r_2$ to get
\begin{equation}
    P(\br_1, \br_2) = F_0^{-1} e^{\frac{1}{2} r_1 \bar r_2 - \frac{1}{4}(r_1 \bar r_1 + r_2 \bar r_2)} \int_{\rm BZ} \frac{d^2 \bk}{2\pi} |\sigma(i k)|^2 e^{-\frac{1}{2} k \bar k} = e^{\frac{1}{2} r_1 \bar r_2 - \frac{1}{4}(r_1 \bar r_1 + r_2 \bar r_2)}
\end{equation}
Interesting, we can view the wavefunctions $\phi_\bk(\br)$ as coherent states $|r\rangle $ labelled by the complex variable $r$ which satisfy the usual coherent state identities
\begin{equation}
    \langle w| z \rangle = e^{\frac{1}{4} (2\bar w z - |z|^2 - |w|^2)}, \qquad \int d^2 z |z \rangle \langle z| = \mathbbm{1}
\end{equation}

The form of the LLL projection is in fact independent of the periodic boundary conditions. We can derive the same relation on the disk with LLL wavefunctions labelled by angular momentum $n$
\begin{equation}
    \phi_n(\br) = \frac{r^n}{2^{\frac{n}{2}}\sqrt{n!}} e^{-\frac{1}{4} r \bar r}, \quad P(\br_1, \br_2) = \sum_n \phi_n(\br_1) \bar \phi_n(\br_2) = e^{-\frac{1}{4}(r_1 \bar r_1 + r_2 \bar r_2)} \sum_n \frac{(r_1 \bar r_2)^n}{2^n n!} = e^{\frac{1}{2} r_1 \bar r_2 - \frac{1}{4}(r_1 \bar r_1 + r_2 \bar r_2)}
\end{equation}

Notice that the action of the projector sends LLL wavefunctions to LLL wavefunctions due to the Bargmann identity. Below we present a simple derivation of this fact that is useful to understand other LLL projectiors. Consider any LLL wavefunction $\psi(z) = f(z) e^{-\frac{1}{4} |z|^2}$.
\begin{align}
    [P \psi](w) &= \frac{1}{2\pi} \int d^2 z P(w, z) f(z) e^{-\frac{1}{4} |z|^2} = e^{-\frac{1}{4}|w|^2} \int d^2 z f(z) e^{\frac{1}{2} \bar z (w - z)} = \frac{1}{2\pi} e^{-\frac{1}{4}|w|^2} \int d^2 z f(z) \frac{2}{w - z} \partial_{\bar z} e^{\frac{1}{2} \bar z (w - z)} \nonumber \\
    &= e^{-\frac{1}{4}|w|^2} \int d^2 z f(z) e^{\frac{1}{2} \bar z (w - z)} \delta(z - w) = f(w) e^{-\frac{1}{4}|w|^2} = \psi(w)
\end{align}
Let us now consider the case where $f(z) = g(z)/z$ meromorphic function with a first order pole (the case of multiple poles can be considered very similarly)
\begin{align}
    [P \psi](w) &= \frac{1}{2\pi} \int d^2 z P(w, z) \frac{g(z)}{z} e^{-\frac{1}{4} |z|^2} = e^{-\frac{1}{4}|w|^2} \int d^2 z \frac{g(z)}{z} e^{\frac{1}{2} \bar z (w - z)} = \frac{1}{2\pi} e^{-\frac{1}{4}|w|^2} \int d^2 z \frac{g(z)}{z} \frac{2}{w - z} \partial_{\bar z} e^{\frac{1}{2} \bar z (w - z)} \nonumber \\
    &= e^{-\frac{1}{4}|w|^2} \int d^2 z g(z) e^{\frac{1}{2} \bar z (w - z)} [\frac{1}{z} \delta(z - w) - \frac{1}{w - z} \delta(z)] = \frac{e^{-\frac{1}{4}|w|^2}}{w} [g(w)  -  g(0)]
\end{align}
Importantly, the pole is cancelled after the LLL projection. Finally we consider the LLL projection of the function $f(z, \bar z) = e^{\frac{1}{2} \bar z \xi} g(z)$:
\begin{align}
    [P \psi](w) &= \frac{1}{2\pi} \int d^2 z P(w, z) g(z) e^{\frac{1}{2} \xi \bar z} e^{-\frac{1}{4} |z|^2} = e^{-\frac{1}{4}|w|^2} \int d^2 z g(z) e^{\frac{1}{2} \bar z (w - z + \xi)}  \nonumber \\
    &= e^{-\frac{1}{4}|w|^2} \int d^2 z g(z) e^{\frac{1}{2} \bar z (w - z + \xi)} \delta(w - z + \xi) = e^{-\frac{1}{4}|w|^2} g(w + \xi)
\end{align}
This last identity implements the fact that the LLL projection of $\bar z$ is $2 \partial_z$. The results for the action of the projector are summarized in Table \ref{tab:LLLProjection}.

\begin{table}[]
    \centering
    \begin{tabular}{c|c}
    \hline \hline
        $\psi$ & $P \psi$ \\
        \hline
        $g(z) e^{-\frac{1}{4} |z|^2}$ & $g(z) e^{-\frac{1}{4} |z|^2}$ \\
        $\frac{g(z)}{z} e^{-\frac{1}{4} |z|^2}$ & $\frac{1}{z}[g(z) - g(0)] e^{-\frac{1}{4} |z|^2}$ \\
        $g(z) e^{\frac{1}{2} \xi \bar z} e^{-\frac{1}{4} |z|^2}$ & $g(z + \xi) e^{-\frac{1}{4} |z|^2}$ \\
        \hline \hline
    \end{tabular}
    \caption{Summary of LLL projectors}
    \label{tab:LLLProjection}
\end{table}

\section{Diagrammatic method for the wavefunction overlaps}
We will now describe a method to compute the the overlap of two wavefunctions $\langle \psi_1|\psi_2 \rangle$ where $|\psi_{1,2} \rangle$ has the form
\begin{equation}
    |\psi_{1,2}\rangle = \A g_{1,2}(\bk^h_{n+1}) \prod_{l=1}^n f_{1,2}(\bk^h_l, \bk^e_l)
\end{equation}
All the overlaps needed to compute the variational energy in this work can be cast in this form. First, consider the normalization for the wavefunctions $|\psi_\bQ \rangle$ for a fixed total momentum $\bQ$. Although the geminal form of the wavefunction (\ref{PsiNGeminal}) is a superposition of all different total momentum states, we can impose a total momentum when computing the expectation values by writing 
\begin{align}
    \langle \psi_\bQ |\psi_\bQ \rangle &= \sum_\bG e^{i \bG \wedge (\bQ - \sum_{l=1}^{n+1} \bk_l^h + \sum_{l=1}^n \bk^l_e)} \frac{\langle \psi_\bxi|\psi_\bxi \rangle}{|\phi_\bQ(\bxi)|^2} \nonumber \\
    &= \sum_\bG e^{i \bG \wedge \bQ} \frac{\langle \psi_\bxi|\psi^\bG_\bxi \rangle}{|\phi_\bQ(\bxi)|^2}
\end{align}
where $\bG$ is a sum over reciprocal lattice vectors. Here, $\psi_{\bxi}^\bG$ has the same form (\ref{PsiNGeminal}) with the functions $g_\bxi$ modified to
\begin{equation}
    g_\bxi(\bk) \mapsto g^\bG_{\bxi}(\br) = e^{-i \bG \wedge \bk} g_\bxi(\bk)
\end{equation}
In the LLL, the polaron dispersion is perfectly flat due to continuous magnetic translation which means that we do not need to impose momentum conservation and can directly work with the states $|\psi_\bxi \rangle$ which are superpositions of different total momenta.

The expectation value of the interaction can also be expressed terms with the form $\langle \psi_1|\psi_2 \rangle$ as follows. In general, the Hamiltonian contains two terms: a single particle term and one that has the form of a sum of terms of the form $\hat O_\bq^\dagger \hat O_\bq$ where $\hat O_\bq$ is a single particle term (cf.~Eq.~\ref{VFirstQuantized}). For the single particle term, we notice that the expectation value of any operator of the form $\hat H_{\rm sp} = \sum_i h_{\sigma_i}(\bk_i)$ can be evaluated for wavefunction of the form (\ref{PsiNGeminal}) by defining
\begin{gather}
    g_\varepsilon(\bk) = [1 + \varepsilon h_h(\bk)] g(\bk),\\
    f_\varepsilon(\bk^h, \bk^e) = [1 + \varepsilon (h_h(\bk^h) + h_e(\bk^e))] f(\bk^h, \bk^e)
\end{gather}
where $\varepsilon$ is some real number. Defining $|\psi_\varepsilon\rangle = \A g_\varepsilon(\bk^h_{n+1}) \prod_{l=1}^n f_\varepsilon(\bk^h_l, \bk^e_l)$,
it is straightforward to see that the expectation value of $\hat H_{\rm sp}$ is
\begin{equation}
    \langle \psi|\hat H_{\rm sp}|\psi \rangle = \frac{d}{d\varepsilon} \langle \psi|\psi_\varepsilon \rangle |_{\varepsilon = 0}
\end{equation}
which is easy to numerically by doing a linear fit at small $\varepsilon$. For the two-body part of the interactions, we define
\begin{gather}
    g_{\bq,\varepsilon}(\bk) = [1 + \varepsilon T^h_\bq]g(\bk)\\
    f_{\bq,\varepsilon}(\bk^e, \bk^h) = [1 + \varepsilon (T^e_\bq - T^h_{-\bq})] f(\bk^e, \bk^h)
\end{gather}
where $\varepsilon$ is complex. This allows us to write
\begin{equation}
    \langle \psi| \hat O_\bq^\dagger \hat O_\bq |\psi \rangle = \frac{\partial^2}{\partial \varepsilon \partial \varepsilon^*} \langle \psi_{\bq,\varepsilon}|\psi_{\bq,\varepsilon} \rangle|_{\varepsilon \rightarrow 0}
\end{equation}
The right hand side can be evaluated numerically using a quadratic fit in the real and imaginary parts of $\varepsilon$.

\begin{figure}
    \centering
    \includegraphics[width = 0.48 \textwidth]{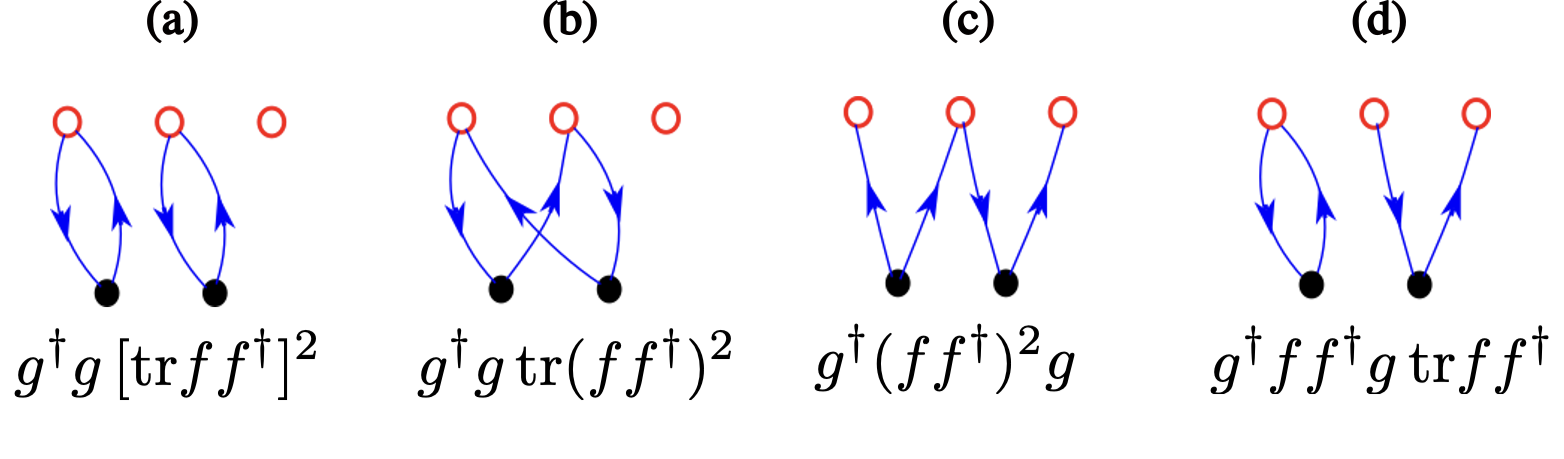}
    \caption{Schematic illustration of the different diagrams contributing to the evaluation of the normalization integral for a two-spin-flip polaron ($n = 2$).}
    \label{fig:diagrams_n_2}
\end{figure}

We will now described how to compute the wavefunction overlap $\langle \psi|\tilde \psi \rangle$ where $|\psi \rangle$ and $|\tilde \psi \rangle$ has the geminal form (\ref{PsiNGeminal})
The overlap integral takes the form $\int \prod_{l=1}^{n+1} d^2 \bk_h^e \prod_{l=1}^n d^2 \bk_l^e \psi^* \tilde \psi$, it suffices to include the antisymmetrization in only one of the functions since $\langle \A u|\A v \rangle = \langle u|\A v \rangle$. Thus, the overlap integral invovles $n! (n+1)!$ terms. Furthermore, it is sufficient to do the antisymmetrization in the hole momenta leading to $(n+1)!$ terms. It turns out that these terms have a simple graphical representation which can be illustrated by the example of $n=2$. In this case, the overlap integral is given by
\begin{equation}
    \langle \psi|\tilde \psi \rangle \propto \int g^*_3 f^*_{11} f^*_{22} [ (\tilde f_{11} \tilde f_{22} - \tilde f_{21} \tilde f_{12}) \tilde g_3
    - (\tilde f_{11} \tilde f_{32} - \tilde f_{31} \tilde f_{12}) \tilde g_2 +  (\tilde f_{21} \tilde f_{32} - \tilde f_{31} \tilde f_{22}) \tilde g_1] 
\end{equation}
where we introduced the shorthand notation $f_{ij} := f(\bk^h_i, \bk^e_j)$, $g_i = g(\bk^h_i)$ and dropped the integration measure for simplicity. We can think of $f_{ij}$ as a matrix in the indices $i$ and $j$ and $g_i$ as a vector and replace the integration with a summation. Defining the matrix $F = \tilde f f^\dagger$, the expression above can be written more compactly as
\begin{equation}
    \langle \psi|\tilde \psi \rangle \propto [\tr F]^2 g^\dagger \tilde g - \tr (F^2) g^\dagger \tilde g - 2g^\dagger F \tilde g \tr F\\ + 2g^\dagger F^2 g 
\end{equation}
This expression only depends on the objects
\begin{equation}
    A_l = \tr F^l, \qquad B_l = g^\dagger F^l g
    \label{AlBl}
\end{equation}
with $l \leq n$, which can be efficiently computed using standard matrix computation techniques. To understand the structure of such expressions for general $n$, we can pictorially represent the $n+1$ hole momenta by an upper horizontal row of $n+1$ dots and the $n$ electrom momenta by a lower horizontal row of $n$ dots as shown in Fig.~\ref{fig:diagrams_n_2}. We denote factors of $f^*$ by directed lines from the upper row to the lower row and factors of $\tilde f$ by directed lines from the lower row to the upper row. We consider all ways of connecting these dots with the condition that each dot has at most one line in and one line out of it. For $n = 2$, this yields the four diagrams in Fig.~\ref{fig:diagrams_n_2}. We see that every closed loop with $n$ electron (lower) nodes corresponds to a factor $\tr F^n$ and  every open loop with $n$ electron momenta corresponds to $g^\dagger F^n g$ (with the unconnected hole dot corresponding to $g^\dagger g$).

Let us first focus on diagrams with one dot unconnected to the rest (corresponding to the factor $g^\dagger g$). We see that these diagrams consists of a number of closed loops whose sizes add up to $n$. This means that these diagrams are in one-to-one correspondence with the integer partitions of $n$. We specify a partition $P$ of $n$ containing $N(P)$ elements by two sets of positive integers $m(P) = \{m_l(P)\}$ and $c(P) = \{c_l(P)\}$ representing the distinct elements of the parition and their count, respectively, such that $n = \sum_{l=1}^{N(P)} c_l(P) m_l(P)$. The value of such diagram for a given partition $P$ will be $B_0 \prod_{l=1}^{N(P)} [A_{m_l(P)}]^{c_l(P)}$. There are two remaining factors to compute the contribution to the overlap integral. One is the sign of this term which can be obtained by noting that any loop with $n+1$ electron nodes requires $n$ exchanges which means that the diagram associated with a partition $P$ will contribute to the overlap integral with the sign $\prod_{l=1}^{N(P)} (-1)^{c_l(P) [m_l(P) - 1]}$. Finally, there is a combinatorial factor that can be understood as the number of permutations of $n$ with cycles whose length are given by the partition $P$ (this is called a cycle type or cycle structure). This number is given by $\frac{n!}{\prod_{l=1}^{N(P)} c_l(P)! m_l(P)^{c_l(P)}}$. The second class of diagrams (diagrams (c) and (d) in Fig.~\ref{fig:diagrams_n_2}) is obtained by opening up one of the closed loops of the first class and connecting it to the lone hole node. Thus, for every partition $P$, we get $N(P)$ diagrams corresponding to opening up each of the distinct loops. The $r$-th diagram will contribute $c_r(P) m_r(P) B_{m_r(P)} \prod_{l=1}^{N(P)} [A_{m_l(P)}]^{c_l(P) - \delta_{l,r}}$ with an opposite sign to the `parent' closed loop diagram. The results can be summarized in the expression
\begin{equation}
    \langle \psi|\tilde \psi \rangle = \sum_P s_P \lambda_P \left\{B_0 \prod_{l=1}^{N(P)} [A_{m_l(P)}]^{c_l(P)}  - \sum_{r=1}^{N(P)} c_r(P) m_r(P) B_{m_r(P)} \prod_{l=1}^{N(P)} [A_{m_l(P)}]^{c_l(P) - \delta_{l,r}}\right\}
    \label{OverlapFinal}
\end{equation}
with $A_n$ and $B_n$ defined in (\ref{AlBl}) and $s_P$ and $\lambda_P$ given by
\begin{equation}
    s_P = \prod_{l=1}^{N(P)} (-1)^{c_l(P) [m_l(P) - 1]}, \quad \lambda_P = \frac{n!}{\prod_{l=1}^{N(P)} c_l(P)! m_l(P)^{c_l(P)}}
\end{equation}
Once we have pre-computed $A_l$ and $B_l$, the evaluation of each term in (\ref{OverlapFinal}) is $O(1)$. This means that, although the number of partitions grows as $\sim \frac{1}{4n \sqrt{3}} e^{\pi \sqrt{\frac{2n}{3}}}$ for large $n$, this scaling is not the limiting factor for the computation up to $n \approx 20$ where the number of partitions is $\sim 600$. The reason is that, as expected, polarons with more spin flips $n$ require a larger system size to estimate their energies accurately. In practice, for system sizes we can investigate (up to $21 \times 21$), the computation of $A_l$ and $B_l$ (see Eqs.~\ref{AlBl}) turns out to be the limiting factor rather than the scaling of the number of diagrams. 

\end{document}